 \documentclass[final,5p,times,twocolumn,authoryear]{elsarticle}

\usepackage{amssymb}
\usepackage{lipsum}
\usepackage{xcolor}
\usepackage{amsmath}
\setcitestyle{square,numbers,sort&compress}
\setcitestyle{citesep={,}}
\biboptions{numbers,sort&compress}

\makeatletter
\def\ps@pprintTitle{%
	\let\@oddhead\@empty
	\let\@evenhead\@empty
	\let\@oddfoot\@empty
	\let\@evenfoot\@oddfoot
}
\makeatother

\usepackage{caption}
\usepackage{float}
\usepackage{graphicx}
\usepackage{subcaption}

\begin{document}

\begin{frontmatter}



\title{Holographic Einstein Ring of AdS-Reissner–Nordström Black Holes with Euler–Heisenberg Nonlinear Electrodynamics}


\author[first]{Abhishek Baruah\fnref{1}}
\fntext[1]{Email: rs\_abhishekbaruah@dibru.ac.in }

\affiliation[first]{organization={Department of Physics, Dibrugarh University},
            city={Dibrugarh},
            postcode={786004}, 
             state={Assam},
            country={India}}

\author[second]{Prabwal Phukon${^{a,}}$\fnref{2}}
\fntext[2]{Email: prabwal@dibru.ac.in (Corresponding author)}
\affiliation[second]{organization={Theoretical Physics Division, Centre for Atmospheric Studies, Dibrugarh University},
            city={Dibrugarh},
            postcode={786004}, 
             state={Assam},
            country={India}}

\begin{abstract}
This study, situated within the framework of the AdS/CFT correspondence, employs wave optics methods to investigate the Einstein ring structure of quantum-corrected AdS–Reissner–Nordstr$\ddot{o}$m black holes governed by Euler–Heisenberg nonlinear electrodynamics. A wave source placed on the AdS boundary yields a response function on the antipodal side, from which a virtual optical system with a convex lens reconstructs the holographic image of the Einstein ring. The analysis systematically explores the impact of physical parameters and observer position on the ring's morphology. As the observer's position varies, the image transitions from a complete ring to an arc and eventually to a single bright point. The Einstein ring radius is observed to decrease with increasing radial source position $\rho$, wave frequency $\omega$, and chemical potential $\mu$, while it increases with electric charge $e$ and temperature $T$. In contrast, the quantum correction parameter $a$ has negligible effect on the ring radius or response amplitude, as its contribution falls off rapidly near the boundary and remains subleading in the wave dynamics. The parameter $e$ enhances the electromagnetic lensing strength, leading to a broader ring, whereas increasing $\rho$ alters wavefront propagation, affecting both brightness peak and ring location. Geometric optics analysis confirms that the incident angle of the photon ring matches the Einstein ring angle, validating consistency across frameworks. Overall, the results highlight how nonlinear electromagnetic effects and bulk field configurations manifest in observable boundary features, providing a means to distinguish quantum-corrected black holes from classical solutions.

\end{abstract}




\end{frontmatter}




\section{Introduction}
\label{introduction}

Since the introduction of the AdS/CFT correspondence by Maldacena \cite{Maldacena:1997re}, the theoretical underpinnings and diverse applications of this duality have generated profound interest and rigorous discourse among physicists. The correspondence not only confirms the theoretical coherence of the framework but also establishes a powerful new method for investigating strongly coupled quantum field theories. Due to their intrinsic complexity, conventional analytical techniques often prove insufficient for uncovering the fundamental properties of these theories directly. In contrast, the AdS/CFT correspondence provides an innovative and robust methodology that allows physicists to probe the complex dynamics of quantum field theories through the advanced tools and perspectives offered by gravitational theories.
\\
In recent years, the AdS/CFT correspondence has found extensive applications across a broad spectrum of physical sciences. Within the domain of condensed matter physics, investigations into holographic superconductors \cite{Hartnoll:2008kx,Hartnoll:2008vx,Gubser:2008px} have unveiled novel theoretical insights into the underlying mechanisms of superconductivity. Furthermore, studies focused on Fermi and non-Fermi liquid behavior \cite{Cubrovic:2009ye,Liu:2009dm,Faulkner:2009wj} have significantly advanced our comprehension of the complex characteristics exhibited by these systems. In addition, the correspondence has proven remarkably effective in elucidating phenomena related to quantum phase transitions, particularly in the context of metal-insulator transitions \cite{Mefford:2014gia,Kiritsis:2015oxa,Donos:2014uba}, thereby showcasing its strong predictive power and explanatory depth.
\\
In addition, the AdS/CFT correspondence has significantly contributed to advancements in black hole (BH) physics. Among the most prominent observational features of BHs, their shadows have garnered sustained attention in both theoretical and observational astrophysics \cite{Shipley:2016omi, Bambi:2008jg, Atamurotov:2013dpa, Mishra:2019trb, Nedkova:2013msa, Chen:2024wtw, He:2024yeg, Shaikh:2018lcc, Wang:2025ihg, Yang:2024nin, He:2024amh, He:2024qka, He:2022yse, He:2022aox, He:2021htq, Zeng:2021dlj, He:2025rjq, Panah:2025jheap,Jafarzade:2024cqg,Hendi:2023jcap,Panah:2020npb,Arenas:2023jhep}. A landmark achievement in this domain was accomplished by the Event Horizon Telescope (EHT), which successfully captured the first-ever images of BH shadows corresponding to the supermassive BHs situated at the centers of the M87 galaxy and our Milky Way \cite{EventHorizonTelescope:2019dse, EventHorizonTelescope:2022wkp}, marking a pivotal moment in observational astronomy.
More recently, substantial progress in the study of BH shadows has been made by Hashimoto and collaborators. By incorporating the AdS/CFT framework along with wave optics formalism, they introduced an innovative method to analyze BH shadows and successfully reconstructed the Einstein ring embedded within the shadow structure \cite{Hashimoto:2019jmw}. In particular, investigations focusing on AdS black holes \cite{Hashimoto:2018okj} have demonstrated that holographic BH images can be systematically generated through gravitational lensing response functions, with the Einstein ring distinctly manifested in these visual reconstructions.
This technique has further been extended to the context of holographic superconductor models \cite{Kaku:2021xqp}, where it revealed abrupt, discontinuous variations in the radius of the photon ring. In parallel, studies concerning charged BH configurations \cite{Liu:2022cev} have indicated that while the chemical potential does not influence the radius of the Einstein ring, the temperature dependence introduces unique and non-trivial behavior. Moreover, within the realm of modified gravity theories, numerous intriguing findings have emerged regarding the holographic Einstein rings of BHs \cite{Zeng:2023zlf, Zeng:2023tjb, Zeng:2023ihy, Hu:2023mai, Li:2024mdd, He:2024bll, Luo:2024wih, Zeng:2024ptv, Zeng:2024ybj, Gui:2024owz, He:2024mal}, further underscoring the versatility and depth of the AdS/CFT approach in probing gravitational phenomena from a holographic perspective.\\
One of the enduring mysteries in General Relativity (GR) pertains to comprehending the nature of singularities concealed within black holes (BHs). Likewise, classical electrodynamics as governed by Maxwell's equations also encounters singularities, leading to ultraviolet divergences associated with the self-energy of point charges. To address the problem of infinite self-energy inherent in classical electrodynamics, Born and Infeld introduced nonlinear electrodynamics (NLED), extending beyond traditional Maxwell electrodynamics \cite{32}. Subsequently, Euler and Heisenberg developed an alternative nonlinear electrodynamics known as Euler–Heisenberg (EH) theory, specifically designed to resolve the infinite self-energy issue of charged particles \cite{33}. This theoretical innovation subsequently became significantly influential within string theory research \cite{34,35}.
\\
Euler-Heisenberg electrodynamics stands as a cornerstone theory within nonlinear electrodynamics, effectively characterizing electromagnetic fields under intense electric and magnetic environments, while quantitatively integrating quantum electrodynamic (QED) phenomena. Particularly noteworthy is that when electric and magnetic fields approach critical magnitudes—namely the electric field critical limit ($E_{cr} \approx 10^{18}$ V/m) and magnetic field critical limit ($B_{cr} \approx 10^{10}$ T) significant QED effects such as photon-photon scattering \cite{37,38} and vacuum birefringence \cite{39} emerge. The EH Lagrangian explicitly describes these nonlinear electromagnetic interactions originating from closed electron loops \cite{40}. Historically, EH electrodynamics has effectively described various electromagnetic phenomena, including photon-photon interactions \cite{41,42} and photon splitting in intense magnetic fields \cite{43,44}.
\\
Nonlinear electrodynamics has further proven influential in studying black hole shadows, a topic of considerable astronomical and theoretical interest \cite{45}. Under nonlinear conditions, the propagation of light resembles electromagnetic waves traveling through a classically dispersive medium \cite{46}. In this scenario, photons no longer follow null geodesics in Minkowski spacetime but instead propagate within an effective geometric framework \cite{47}. Novello et al. investigated the geometric properties of light propagation within EH nonlinear electrodynamics in the context of regular black holes, exploring this effective geometry \cite{36}. With advancements in nonlinear quantum electrodynamics, significant research has been directed towards understanding the geometry of photon propagation in nonlinear electrodynamic frameworks \cite{48,49}, as well as in generalized Born-Infeld theories \cite{50}.
The Einstein–Euler–Heisenberg (EEH) model, an exact coupling solution of EH electrodynamics with General Relativity, has been examined extensively across various aspects, including thermodynamics \cite{51}, thermal fluctuations \cite{52}, gravitational lensing phenomena \cite{53}, accretion disk physics \cite{54}, and properties related to black hole shadows and quasinormal modes \cite{55}. 
\\
The structure of this work is organized as follows. In Section 2, we present a concise introduction to Euler–Heisenberg AdS black holes and derive the lensing response function, which characterizes the diffraction behavior of wave sources influenced by the black hole geometry. This function plays a pivotal role in elucidating the gravitational lensing properties inherent to such spacetimes. Section 3 details the experimental framework, wherein an optical system incorporating a convex lens is employed to visualize the formation of Einstein rings on a detection screen. This section further explores the influence of key physical parameters—namely, the quantum correction coefficient $a$ associated with one-loop QED effects in Euler–Heisenberg electrodynamics, the black hole’s electric charge $e$, chemical potential $\mu$ and the observer’s spatial location $\rho$ on the structure and behavior of the resulting Einstein ring. Moreover, a comparative analysis between the predictions obtained from wave optics and those derived from geometric optics is undertaken, underscoring both their convergences and distinctive features. Lastly, Section 4 provides a comprehensive summary of the principal results and conclusions drawn from the study.
\section{Motivations}
 This present work offers a novel investigation of holographic Einstein rings in the background of asymptotically AdS black holes coupled to Euler–Heisenberg electrodynamics, using both wave optics and geometric optics techniques. By computing the lensing response function and analyzing the resulting intensity profiles on a holographic detection screen, we systematically explore how physical parameters—including the quantum correction parameter $a$, scalar field electric charge $e$, chemical potential $\mu$, and observer location 
$\rho$ affect the structure, size, and discontinuities in the Einstein ring. Our study thereby extends the ongoing program of using holographic imaging as a nontrivial probe of modified gravity and nonlinear electrodynamics in the AdS/CFT framework. \\
Furthermore, this analysis sheds light on the interplay between wave-based and ray-based descriptions of light propagation in strongly curved spacetimes. By comparing ring features derived from gravitational lensing response functions with those predicted via geometric photon orbit methods, we critically assess the validity and limitations of geometric approximations in quantum-corrected NLED regimes. Ultimately, this work not only broadens the theoretical understanding of Einstein rings and photon spheres in nonlinear electrodynamic black holes, but also underscores the significant role of quantum corrections in shaping observable features of holographic spacetimes—a question of deep interest in both fundamental gravity and quantum optics.
\section{The Euler-Heisenberg AdS black hole its response
function}
The Euler-Heisenberg AdS black hole, is described by the following spherically symmetric metric: 
\begin{equation}
\mathrm{d}s^{2} =-f(r)\mathrm{d}t^{2}  +\frac{1}{f(r)} \mathrm{d}r^{2} +r^{2} (\mathrm{d}\theta ^{2}+\sin ^{2}\theta \mathrm{d}\phi ^{2}  ),\label{1}
\end{equation}
where the horizon function is\cite{56}.
\begin{equation}
f(r)=1-\frac{2M}{r} +\frac{Q^2}{r^2}+\frac{r^2}{l^2}-\frac{aQ^4}{20r^6}\label{2}
\end{equation}
The parameter $M$ corresponds to the mass of the black hole, serving as a measure of the gravitational strength of the central object. The symbol $l$ denotes the AdS curvature radius, which characterizes the length scale associated with the asymptotic anti-de Sitter (AdS) nature of the spacetime. For the sake of analytical convenience in subsequent derivations, we set $l=1$ without loss of generality. The parameter $a$ encapsulates the effects of quantum corrections arising from one-loop quantum electrodynamics, as introduced via the Euler–Heisenberg nonlinear electrodynamics framework. Meanwhile, $Q$ represents the electric charge of the black hole and governs the electromagnetic contribution to the spacetime geometry. To facilitate the analysis, a change of variables is employed whereby the radial coordinate is transformed as $r=1/z$, and the metric function is redefined accordingly as 
$F(z)=f(1/r)$. Under this transformation, the governing equations originally expressed in Eq.(\ref{1}) and Eq.(\ref{2}) can be reformulated in terms of the new variable $z$ as
 \begin{equation}
\mathrm{d}s^{2} =\frac{1}{z^{2} }[-F(z)\mathrm{d}t ^{2}+\frac{\mathrm{d}z^{2}}{F(z)}+ \mathrm{d}\theta ^{2}+\sin ^{2}\theta \mathrm{d}\varphi ^{2}  ],\label{3}
\end{equation}
\begin{equation}
F(z)=-\frac{1}{20} a Q^4 z^6+\frac{1}{l^2 z^2}-2 M z+Q^2 z^2+1 .
\end{equation}
In the transformed coordinate framework, the spacetime exhibits a curvature singularity in the limit $z \rightarrow \infty$, indicating the presence of a physical divergence in the geometry. Conversely, the conformal boundary of the anti-de Sitter (AdS) spacetime is situated at $z=0$, serving as the asymptotic region relevant for holographic analyses. The thermodynamic temperature of the black hole is characterized by the Hawking temperature, which is computed via the relation $T=\frac{1}{4 \pi} F'(z_h)$, where $z=z_h$ denotes the location of the event horizon in the transformed coordinate system. It is also worth emphasizing that in the limit $z_h\rightarrow \infty$, the black hole vanishes, and the spacetime reduces to a pure AdS geometry, devoid of any horizon structure.\\
Subsequently, we proceed to investigate the behavior of a massless scalar field propagating in the given gravitational background by employing the covariant form of the Klein–Gordon equation \cite{Hashimoto:2018okj}
 \begin{equation}
D_{\alpha}D^{\alpha}\tilde{\Psi}-{\mathcal{M}}^{2}\tilde{\Psi}=0. 
\end{equation}
Here, the operator $D_\alpha\equiv \Delta_\alpha-ieA_\alpha$  denotes the gauge-covariant derivative, where $\Delta_\alpha$ is the standard covariant derivative compatible with the background metric, and $A_\alpha$ corresponds to the electromagnetic four-potential. The quantity $\tilde{\Psi}$ represents a complex scalar field carrying electric charge $e$, while $\mathcal{M}$ denotes the rest mass of the scalar field. In order to simplify the process of solving the Klein–Gordon equation in the curved black hole background, it proves beneficial to adopt the ingoing Eddington–Finkelstein coordinate system \cite{Liu:2022cev}. This coordinate transformation is particularly useful for capturing the causal structure of the black hole spacetime and for preserving the regularity of physical fields across the event horizon. Within this framework, the metric and associated quantities can be reformulated in terms of the new coordinates, which are introduced as follows
  \begin{equation}
\upsilon \equiv t+z_{*} =t-\int \frac{\mathrm{d}z}{F(z)} 
\end{equation}
Accordingly, the nontrivial background fields present in the bulk spacetime undergo transformation and can be recast into the following explicit form given as
  \begin{equation}
\mathrm{d}s^{2} =\frac{1}{z^{2} }\Big[-F(z)\mathrm{d}\upsilon  ^{2}-2\mathrm{d}z \mathrm{d}\upsilon + \mathrm{d}\theta ^{2}+\sin ^{2}\theta \mathrm{d}\varphi ^{2} \Big]
\end{equation}
\begin{equation}
A_{\alpha}=-A(z)(\mathrm{d}\upsilon)_\alpha,
\end{equation}
 In this framework, the electromagnetic four-potential is given by the expression $A(z)=Q(z-z_h)$ where a gauge transformation has been applied to render the potential regular at the horizon. The chemical potential associated with the dual field theory is identified as $\mu=z_h Q$, arising naturally from the temporal component of the gauge field evaluated at the AdS boundary. For the sake of computational clarity we assign the scalar field mass as 
$\mathcal{M}=-2$, which lies above the Breitenlohner–Freedman bound in AdS spacetime. By redefining the scalar field as $\tilde{\Psi}=z\Psi$, the asymptotic behavior of the rescaled field $\Psi$ in the vicinity of the AdS boundary can be systematically characterized as follows
 \begin{eqnarray}
\Psi(\upsilon ,z,\theta, \varphi  )=J_{\mathcal{K}}(\upsilon ,\theta, \varphi )+\left \langle \mathcal{K} \right \rangle z +K(z^{2} ),
\end{eqnarray}
In accordance with the principles of the holographic correspondence, the quantity $J_{\mathcal{K}}$ is interpreted as the external source coupled to an operator in the boundary quantum field theory. This source term drives perturbations in the dual theory, and the corresponding expectation value of the associated boundary operator, commonly referred to as the response function which is determined through the holographic prescription. The explicit expression for this response function is provided below, following the formulation outlined in Ref.~\cite{Liu:2022cev}
 \begin{eqnarray}
\left\langle \mathcal{K}\right\rangle_{J_\mathcal{K}}= \left\langle \mathcal{K}\right\rangle-(\partial\upsilon-i\mu)J_\mathcal{K},\label{10}
\end{eqnarray}
 In this context, the symbol $\left\langle \mathcal{K}\right\rangle$ denotes the vacuum expectation value of the boundary operator in the absence of an external source term. Following the prescription established in \cite{Klebanov:1999tb}, we adopt a monochromatic, axisymmetric Gaussian wave packet as the form of the incident wave source. This source is localized at the South Pole of the AdS boundary, corresponding to the angular position $(\theta_0=\pi)$, as illustrated in Figure 4 of \cite{Liu:2022cev}. This specific configuration facilitates the analysis of wave propagation and the subsequent formation of holographic features such as Einstein rings within the dual gravitational background as
\begin{eqnarray}
J_{\mathcal{k}}  (\upsilon ,\theta )&=&e^{-i\omega \upsilon }(2\pi\eta ^{2} )^{-1} \mathrm{exp}[-\frac{(\pi -\theta )^{2} }{2\eta ^{2} }] \nonumber\\
&=&e^{-i\omega \upsilon }\sum_{l=0}^{\infty }C_{l0} X_{l0}(\theta ) \label{11} 
\end{eqnarray}
In this formulation, the parameter $\eta$ characterizes the spatial width of the wave packet generated by the Gaussian source. It is assumed that $\eta <<\pi$, indicating that the wave is highly localized near the source point on the AdS boundary. The symbol $\omega$ corresponds to the angular frequency of the incoming wave, governing its temporal oscillation behavior. The function 
$X_{l0}(\theta)$ denotes the standard spherical harmonic associated with azimuthal quantum number $m=0$, which encapsulates the angular dependence of the wave on the spherical boundary. The coefficients $C_{l0}$ represent the expansion weights corresponding to each spherical harmonic mode 
$X_{l0}(\theta)$, and these coefficients can be expressed through the following expansion 
\begin{equation}
C_{l0}=(-1)^{l}\sqrt{\frac{l+1/2}{2\pi} }  \mathrm{exp}\left(-\frac{(l+1/2)^{2}\eta ^{2}  }{2} \right).
\end{equation}
 In accordance with Eq.~(\ref{11}), the associated solution within the bulk spacetime can be formulated as follows
  \begin{equation}
\Psi(\upsilon ,z,\theta)=\sum_{l=0}^{\infty } e^{-i\omega \upsilon }C_{l0}Y_{l}(z)
X_{ln}(\theta),\label{13}
\end{equation}
 The radial wave function $Y_l$ is governed by the following equation of motion, which arises from the separation of variables applied to the underlying field equation in the bulk geometry
 \begin{eqnarray}
z^{2}F(z){Y}''_{l} +(z^{2}{F}'(z)-2zF(z)+2i\omega z^{2}){Y}'_{l} \nonumber\\ +[-2i\omega z-z^{2}l(l+1)]{Y}_{l}=0.\label{yundong}
\end{eqnarray}
In accordance with the AdS/CFT correspondence, the radial function $Y_l$ admits an asymptotic expansion near the AdS boundary. This expansion captures the behavior of the bulk field in the near-boundary region and is expressed as follows
  \begin{equation}
Y_{l} =1 +\left \langle \mathcal{K}\right \rangle_{l}z
+K(z^{2} )\end{equation}
 Consequently, the resulting response function $\langle \mathcal{K} \rangle_{J_\mathcal{K}}$, evaluated in the presence of the external source 
$J_\mathcal{K}$, admits the following series expansion
 \begin{equation}
\left\langle \mathcal{K}\right\rangle_{J_\mathcal{K}}=\sum_{l=0}^{\infty } e^{-i\omega \upsilon }C_{l0}\left\langle \mathcal{K}\right\rangle_{J_\mathcal{K}{l}} X_{l0} (\theta ) \label{xiangying}
\end{equation}
 where
  \begin{equation}
\left\langle \mathcal{K}\right\rangle_{J_\mathcal{K}{l}}=\left\langle \mathcal{K}\right\rangle_{l}+i\tilde{\omega}.
\end{equation}
In this context, the modified frequency is defined as $\tilde{\omega}=\omega+e\mu$, where $\mu$ denotes the chemical potential and $e$ is the charge of the scalar field. The radial function $Y_l$ is subject to two boundary conditions: one imposed at the black hole event horizon, located at $z=z_h$ , and the other at the asymptotic AdS boundary, corresponding to $z=0$. The external wave source $J_\mathcal{K}$ is identified with the asymptotic behavior of the scalar field near the boundary. In particular, the boundary condition $Y_l(0)=1$ is adopted, as inferred from Eq.~(\ref{yundong}).
To determine the radial profile $Y_l$ , we numerically solve the equation of motion using the pseudo-spectral method \cite{Hashimoto:2019jmw,Hashimoto:2018okj}, which provides highly accurate solutions in curved spacetime backgrounds. From the obtained numerical data, the partial response function 
$\langle\mathcal{K}\rangle_l$ is extracted. Subsequently, the total response function is computed using Eq.~(\ref{xiangying}).
It is important to note that, since the observation point is located sufficiently far from the localized wave source, the contribution of the second term associated with the source in Eq.(\ref{10}) becomes negligible. As a result, within the framework of Eq.(\ref{yundong}), the full response 
$\langle\mathcal{K}\rangle_{J_\mathcal{K} l}$ can be effectively approximated by the partial response $\langle \mathcal{K} \rangle$

\begin{figure}
	\centering 
\includegraphics[width=0.4\textwidth, angle=0]{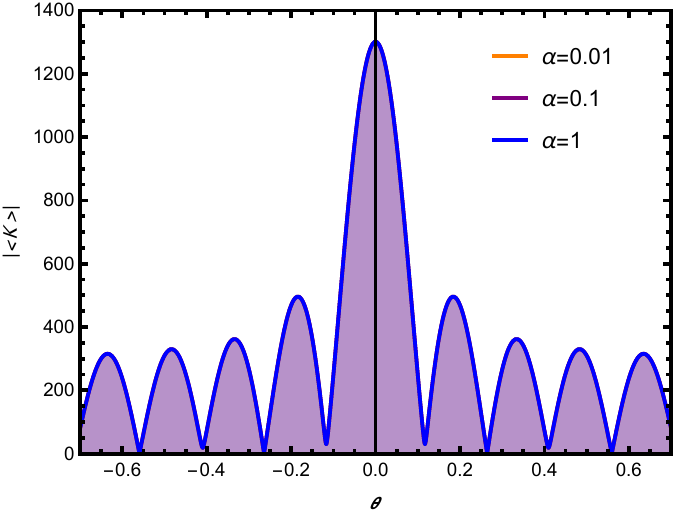}	
	\caption{Response function for different $\alpha$ with $Q=0.5$, $z_{h}=1$, $e=1$, $\omega =90$}
	\label{1}%
\end{figure}

\begin{figure}
	\centering 
\includegraphics[width=0.4\textwidth, angle=0]{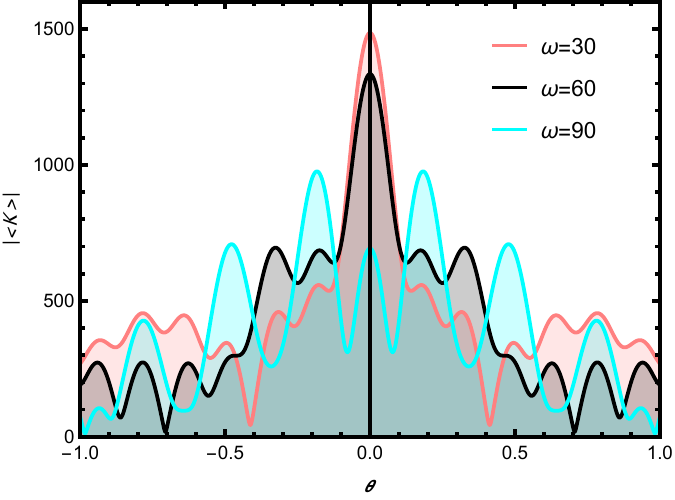}	
	\caption{Response function for different $\omega$ with $a=1$, $Q=0.9$, $y_{h}=1.2$, $e=1$.}
	\label{2}%
\end{figure}

\begin{figure}
	\centering 
\includegraphics[width=0.4\textwidth, angle=0]{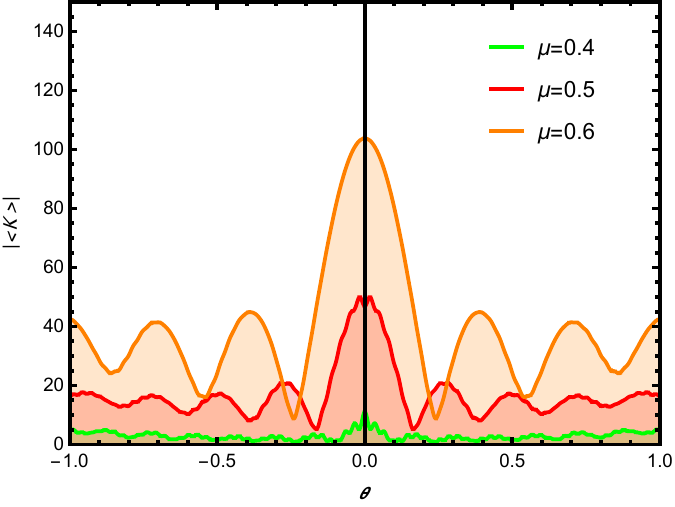}
	\caption{Response function for different $\mu$  with $a=1$, $y_{h}=1$, $e=0.5$, $\omega=90$, $\rho=20$ from top to bottom,
the values of $\mu$ correspond to $Q=0.4, 
0.5, 0.6$, respectively.}
	\label{3}%
\end{figure}

\begin{figure}
	\centering 
\includegraphics[width=0.4\textwidth, angle=0]{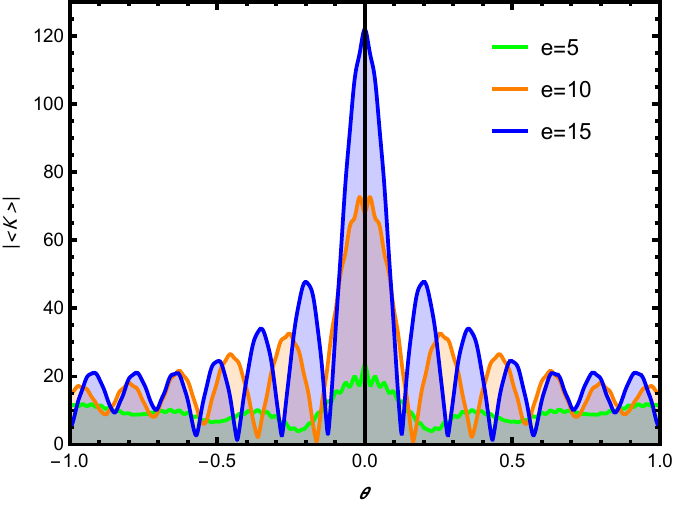}
	\caption{Response function for different $e$ with $a=1$, $Q=0.5$, $z_{h}=1$, $\omega=90$, $\rho=20$.}
	\label{4}%
\end{figure}
\begin{figure}
	\centering 
\includegraphics[width=0.4\textwidth, angle=0]{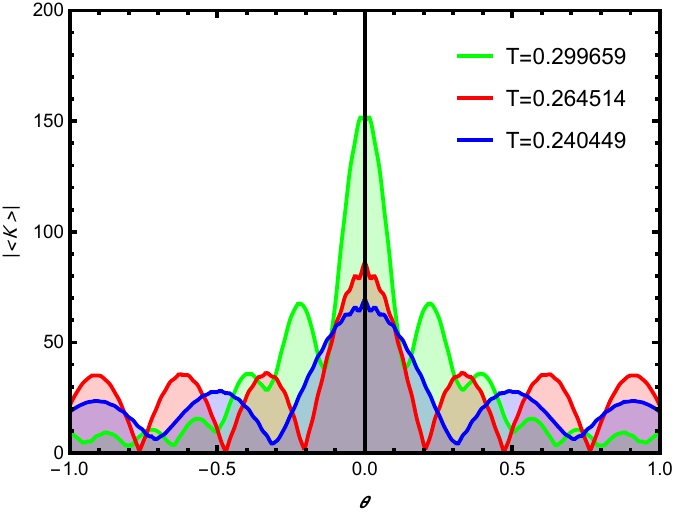}	
	\caption{Response function for different $T$  with $a=0.1$, $e=0.5$, $\omega=90$, $\rho=20$. From top to bottom,
the values of $T$ correspond to $z_{h}=1, 
1.2, 1.4$, respectively.}
	\label{5}%
\end{figure}
\begin{figure}
	\centering 
\includegraphics[width=0.4\textwidth, angle=0]{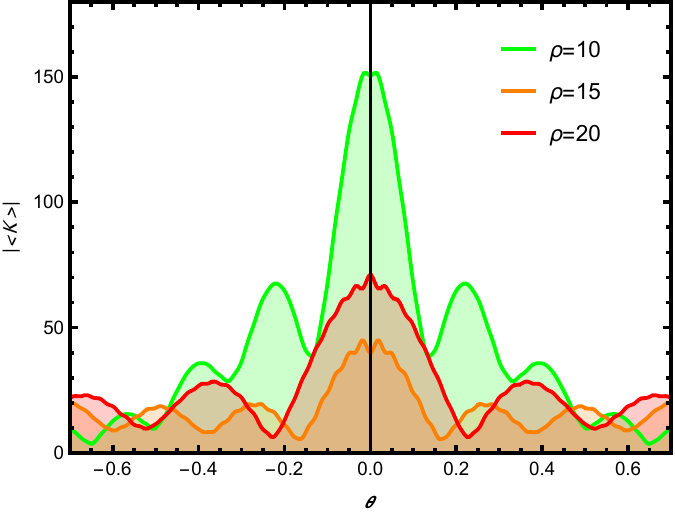}	
	\caption{Response function for different $\rho$  with $a=1$, $z_{h}=1$, $e=0.5$, $\omega=90$. 
}
	\label{6}%
\end{figure}

Figures \ref{1} through \ref{6} collectively depict the behavior of the response function under variations of several key physical parameters. In particular, Figure \ref{1} demonstrates that the magnitude of the response function remains largely unaffected as the quantum correction parameter 
$a$ is increased. In contrast, Figure \ref{2} reveals a diminishing trend in amplitude with increasing values of the incident wave frequency $\omega$. As illustrated in Figure \ref{3}, the response amplitude exhibits an enhancement as the chemical potential $\mu$ becomes more pronounced. Similarly, Figure \ref{4} indicates that a greater scalar field electric charge $e$ leads to a higher response amplitude. Figure \ref{5} shows a monotonic increase in amplitude with rising black hole temperatures $T$. Lastly, Figure \ref{6} highlights an inverse relationship between the response amplitude and the observer’s radial position $\rho$, where larger values of $\rho$ correspond to a suppressed amplitude\\
Following a comprehensive analysis of the response function across these varying parametric regimes, we infer that the response essentially manifests as a diffraction pattern produced by the interaction of an incident wave source with the black hole geometry. To further elucidate the phenomenology associated with black hole shadows, we propose the incorporation of an optical detection apparatus being specifically, a convex lens configuration which is engineered in such a way that it observes the formation of the Einstein ring with greater precision.

\section{The emergence of the Einstein ring structure}
To enable the detection of Einstein rings in the holographic framework, an optical arrangement incorporating a convex lens must be introduced, as described in Ref.~\cite{Liu:2022cev}. This lens plays a crucial role by converting planar wavefronts, emanating from a localized Gaussian source, into spherical waves that propagate toward the boundary of the AdS spacetime. This transformation is vital for capturing diffraction patterns, such as Einstein rings, as perceived from specific angular directions.
In particular, the observational angle is denoted by $\theta_{obs}$, which defines the direction along which the response function is evaluated on the AdS boundary. To accommodate this observational setup, the standard angular coordinates $(\theta, \varphi)$ are rotated to a new coordinate basis that aligns the polar axis with the direction of observation. The transformed coordinate system satisfies a well-defined geometric relationship, which ensures the preservation of rotational symmetry and facilitates the computation of the response at arbitrary angles, as elaborated in Ref.~\cite{Liu:2022cev} and given as
 \begin{equation}
  \cos\varphi ^{'}  +i\cos\theta ^{'}  =e^{i\theta _{obs} }  (\sin\theta  \cos\varphi +i\cos\theta  )
  \end{equation}
 In this optical framework, the angular coordinates $\theta^{'}=0$ and 
 $\varphi^{'}=0$ are chosen to define the central axis of observation. A Cartesian coordinate system $(x,y,z)$ is introduced such that the boundary plane where
  the observer resides is represented by $(x,y)=(\theta^{'}\cos\varphi^{'},\theta^{'} \sin \varphi^{'})$, establishing a virtual observation plane within the AdS boundary geometry.
To simulate optical imaging, a convex lens is positioned within this two-dimensional plane $(x,y)$, acting as an interface that manipulates the propagation of the scalar wave. The lens is characterized by its focal length $f$, while $d$ denotes the corresponding effective aperture radius. A spherical detection screen is placed at a distance $f$ from the lens, and the coordinates on this screen are defined by $(x,y,z)=(x_{SC},y_{SC},z_{SC})$, which satisfy the geometric constraint of a sphere: $x^2_{SC}+y^2_{SC}+z^2_{SC}=f^2$.
The wave dynamics are captured by analyzing the transformation of the scalar field as it passes through the lens. Specifically, the relationship between the incoming scalar wave $\Psi(\vec{x})$ before interacting with the lens, and the outgoing transmitted wave $\Psi_T(\vec{x})$ after lensing, is governed by the following expression
   \begin{equation}
\Psi _{T} (\vec{x} )=e^{-i\tilde{\omega }\frac{\left |\vec{x} \right |^2 }{2f} } \Psi (\vec{x} ).
\end{equation}
 The resulting wave function projected onto the detection screen can be represented by the following expression, as formulated in Ref.~\cite{Liu:2022cev}
  \begin{eqnarray}
\Psi _{SC} (\vec{x}_{SC}  )&=&\int_{\left |\vec{x}  \right |\le d}d^{2} x\Psi _{T} (\vec{x} )
e^{i\tilde{\omega } R}\nonumber\\&\propto &\int_{\left |\vec{x}  \right |\le d}d^{2} x\Psi(\vec{x} )e^{-i\frac{\tilde{\omega }}{f}
  \vec{x}\cdot \vec{x}_{SC} }\nonumber\\&=&\int d^{2} x\Psi (\vec{x} )\sigma (\vec{x})e^{-i\frac{\tilde{\omega }}{f}
  \vec{x}\cdot \vec{x}_{SC} },\label{bo}
\end{eqnarray}
 In this formulation, the variable $R$ denotes the spatial distance between a point $(x,y,0)$ located on the plane of the convex lens and a point $(x^2_{SC},y^2_{SC},z^2_{SC})$ situated on the observation screen. Within the same expression, $\sigma(\vec{x})$ represents the window function, which is defined as follows
 \begin{equation}
\sigma (\vec{x}): = 
\begin{cases}
  1,~~~~0\le\left |\vec{x}  \right | \le d;  \\
  0,~~~~~~~~~~~\left |\vec{x}  \right | >  d.
\end{cases} 
\end{equation}
As demonstrated by Eq.(\ref{bo}), the wave profile detected on the observation screen is mathematically related to the incoming wave via a Fourier transformation. In the present analysis, the response function is interpreted as the incident wave $\Psi(\vec{x})$, allowing holographic projections such as Einstein ring structures to be visualized on the detection screen through the application of Eq.(\ref{bo}).
To systematically explore the influence of key physical parameters on the resulting holographic images, we adopt a fixed source width of $\eta=0.02$, characterizing the narrowness of the Gaussian wave packet, and a convex lens with radius $d=0.6$, which defines the effective aperture of the optical system. These parameters serve as the baseline for investigating how variations in black hole and field properties modulate the observed holographic patterns.

\subsection{Influence of spacetime-structure-dependent parameters on the formation and characteristics of Einstein rings.}

Figure {\ref8} illustrates the influence of the observer’s angular position 
$\theta_{obs}$ on the morphology of the Einstein rings. When $\theta_{obs}=0$, the observer is situated at the North Pole of the AdS boundary. As 
$\theta_{obs}$ increases from 0 to $\pi/6$, $\pi/3$ and subsequently to $\pi/2$, the ring structure progressively diminishes, ultimately leaving only bright spots visible at $\theta_{obs}=\pi/2$ . This behavior indicates a strong dependence of the Einstein ring's geometry on the observer’s location.
 Figure {\ref9} investigates the influence of the parameter $a$ on the structure of Einstein rings, revealing that an increase in $a$ leads to no change in the ring radius. The no change condition is more distinctly visible in Figure {\ref10}, where the x-axis denotes the radial position corresponding to peak intensity interpreted as the Einstein ring radius while the y-axis represents the associated luminosity. Specifically, for $a=0.01,0.1,1,10$, the peak positions are located only at $x_s/f=0.44$. These values clearly demonstrate that the Einstein ring radius remains constant with increasing $a$.

\begin{figure}[htbp]
  \centering
  \begin{subfigure}[b]{0.48\columnwidth}
    \centering
    \includegraphics[width=\textwidth,height=0.8\textwidth]{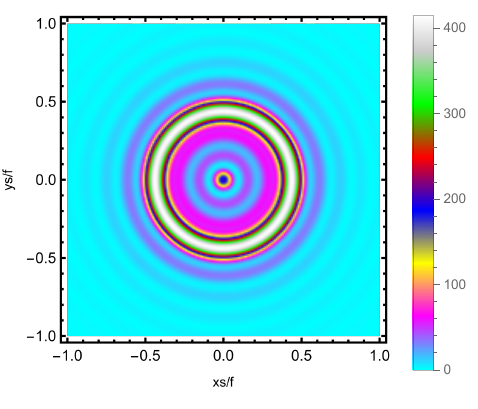}
    \caption{$\theta_{obs}=0$}
  \end{subfigure}
  \hfill
  \begin{subfigure}[b]{0.48\columnwidth}
    \centering
    \includegraphics[width=\textwidth,height=0.8\textwidth]{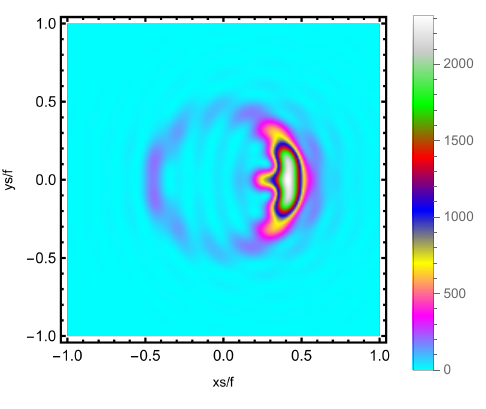}
    \caption{$\theta_{obs}=30$}
  \end{subfigure}
\begin{subfigure}[b]{0.48\columnwidth}
    \centering
    \includegraphics[width=\textwidth,height=0.8\textwidth]{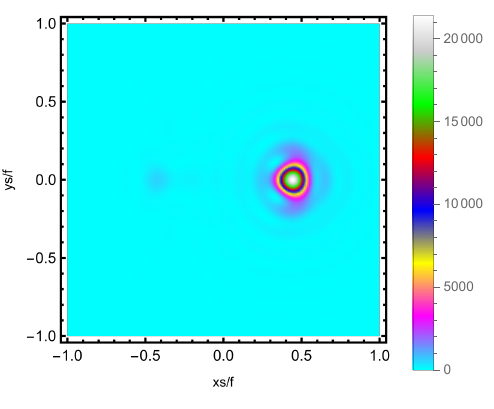}
    \caption{$\theta_{obs}=60$}
  \end{subfigure}
  \hfill
  \begin{subfigure}[b]{0.48\columnwidth}
    \centering
    \includegraphics[height=0.8\textwidth,width=\textwidth]{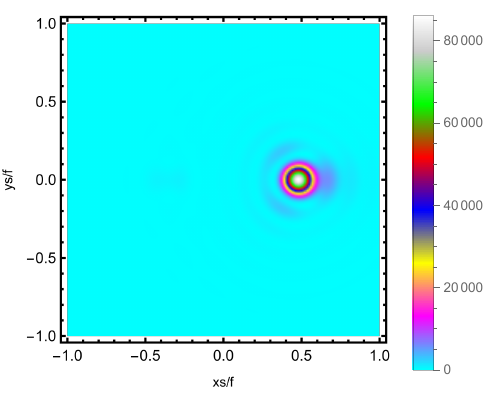}
    \caption{$\theta_{obs}=90$}
  \end{subfigure}
  \caption{Effect of the observer’s angular position $\theta_{obs}$ on the formation of the Einstein ring, with fixed parameter values: $a=0.1$, $Q=0.5$, $e=1$, $z_{h}=1$, $\omega=40$, $\rho=15$.}
  \label{8}%
\end{figure}

\begin{figure}[htbp]
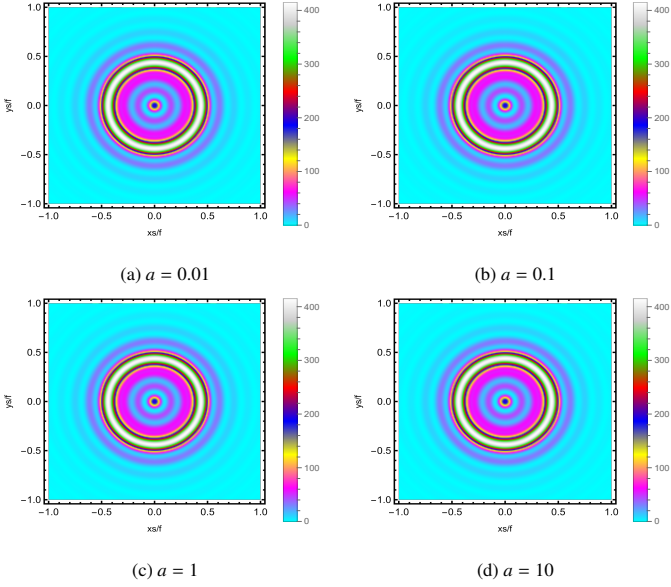

  \centering
  \begin{subfigure}[b]{0.48\columnwidth}
    \centering
    \includegraphics[width=\textwidth,height=0.8\textwidth]{Figure8a.pdf}
    \caption{$a=0.01$}
  \end{subfigure}
  \hfill
  \begin{subfigure}[b]{0.48\columnwidth}
    \centering
    \includegraphics[width=\textwidth,height=0.8\textwidth]{Figure8b.pdf}
    \caption{$a=0.1$}
  \end{subfigure}
\begin{subfigure}[b]{0.48\columnwidth}
    \centering
    \includegraphics[width=\textwidth,height=0.8\textwidth]{Figure8c.pdf}
    \caption{$a=1$}
  \end{subfigure}
  \hfill
  \begin{subfigure}[b]{0.48\columnwidth}
    \centering
    \includegraphics[height=0.8\textwidth,width=\textwidth]{Figure8d.pdf}
    \caption{$a=10$}
  \end{subfigure}
  \caption{Influence of the quantum correction parameter $a$ on the structure and formation of the Einstein ring, where $\theta_{obs}=0$, $Q=0.5$, $e=1$, $z_{h}=1$, $\  \omega=40$, $\rho=15$.}
  \label{9}%
\end{figure}

\begin{figure}[htbp]
  \centering
  \begin{subfigure}[b]{0.48\columnwidth}
    \centering
    \includegraphics[width=\textwidth,height=0.8\textwidth]{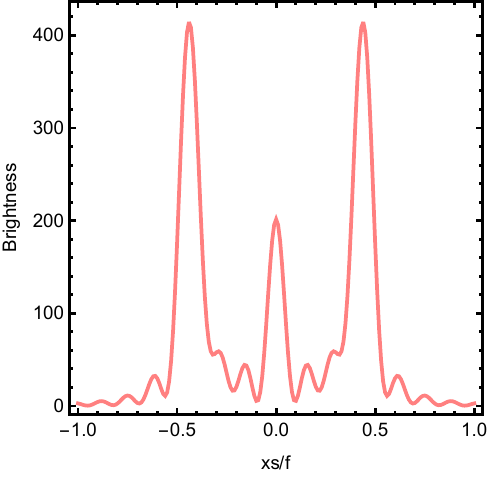}
    \caption{$a=0.01$}
  \end{subfigure}
  \hfill
  \begin{subfigure}[b]{0.48\columnwidth}
    \centering
    \includegraphics[width=\textwidth,height=0.8\textwidth]{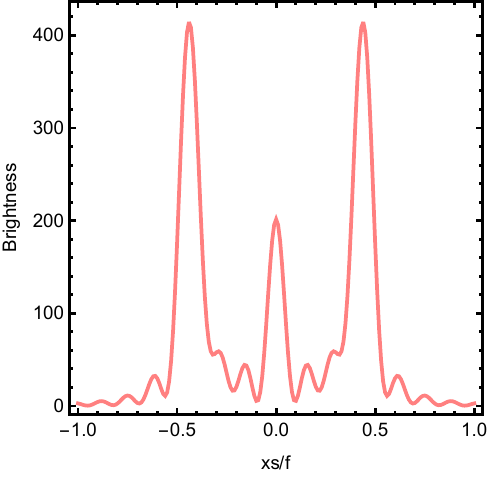}
    \caption{$a=0.1$}
  \end{subfigure}
\begin{subfigure}[b]{0.48\columnwidth}
    \centering
    \includegraphics[width=\textwidth,height=0.8\textwidth]{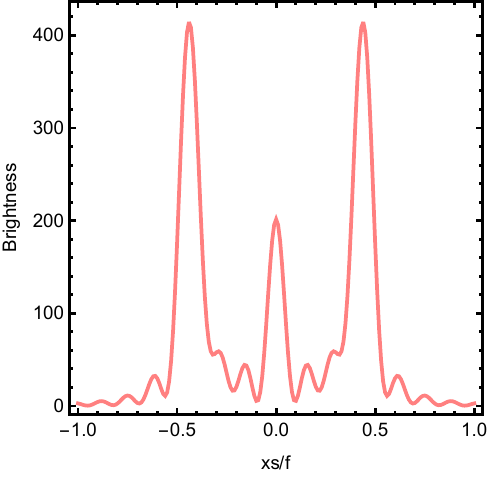}
    \caption{$a=1$}
  \end{subfigure}
  \hfill
  \begin{subfigure}[b]{0.48\columnwidth}
    \centering
    \includegraphics[height=0.8\textwidth,width=\textwidth]{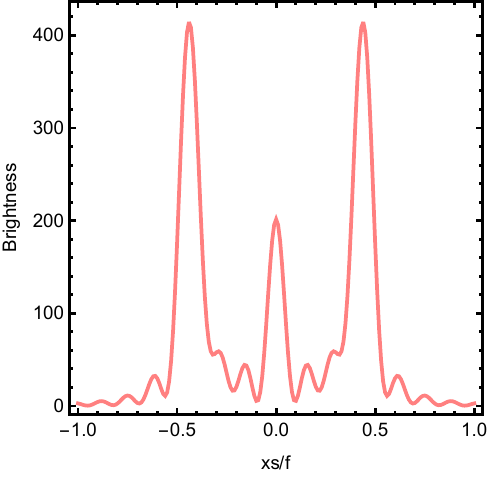}
    \caption{$a=10$}
  \end{subfigure}
  \caption{Influence of the parameter $a$ on the brightness, with the observer’s angle fixed at $\theta_{obs}=0$, and the parameters set as 
$Q=0.5$, $e=1$, $z_{h}=1$, $\  \omega=40$, $\rho=15$}
  \label{10}%
\end{figure}

Similarly, Figure \ref{11} explores the effect of the parameter $e$ which is the electric charge of the scalar field on the Einstein ring structure, demonstrating that an increase in $e$ leads to an expansion in the ring radius. The associated luminosity distribution is presented in Figure \ref{12}, where the brightness reaches its peak at 
$e=20$ and $x_s/f=0.79$. As the value of $e$ increases, the position of the luminosity peaks along the x-axis shifts accordingly, confirming that the Einstein ring radius grows with increasing $e$.
Figure \ref{13} investigates the influence of the parameter $\rho$ on the formation of Einstein rings. Although this effect may not be immediately discernible from the direct visual inspection of the ring profiles, a more detailed assessment complemented by the luminosity distribution shown in Figure~\ref{14} indicating that the Einstein ring radius exhibits a decreasing trend with increasing $\rho$. Notably, when $\rho=10$, the peak luminosity occurs at $x_s/f=0.68$. As $\rho$ increases to 15, 20, and 25, the corresponding peak positions shift to $x_s/f=0.52$, $x_s/f=0.38$, and $x_s/f=0.33$ respectively. These results substantiate a clear inverse relationship between the parameter $\rho$ and the radius of the Einstein ring.
The influence of wave source parameters particularly the wave source frequency $\omega$ on the structure of the Einstein ring is also examined. As illustrated in Figure \ref{15}, a pronounced trend emerges: increasing the frequency $\omega$ leads to a reduction in the radius of the Einstein ring. At lower frequencies, such as $\omega=20$, multiple diffraction fringes are visible, highlighting strong wave interference effects but the resolution is not clear. However, as $\omega$ increases to 40, 60, 80, these secondary diffraction rings progressively diminish, leaving a few, well-defined primary Einstein rings with better resolution. This behavior reflects the growing dominance of geometric optics over wave optics at higher frequencies. To further substantiate this observation, Figure \ref{16} presents the relationship between luminosity and wave source frequency. It is clearly observed that with increasing frequency values $\omega= 20, 40, 60, 80$, the corresponding peak positions shift from $x_s/f=0.71, 0.33, 0.21, 0.14$. This systematic decrease in peak position underscores that higher frequencies yield smaller ring radii, reaffirming the inverse relationship between $\omega$ and the size of the Einstein ring, and confirming the critical role of high-frequency sources in enhancing the ring’s resolution and clarity.

\begin{figure}[htbp]
  \centering
  \begin{subfigure}[b]{0.48\columnwidth}
    \centering
    \includegraphics[width=\textwidth,height=0.8\textwidth]{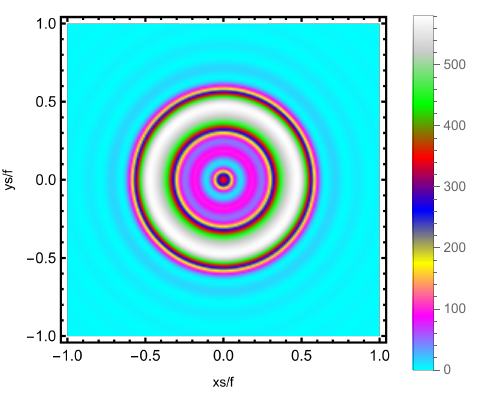}
    \caption{$e=5$}
  \end{subfigure}
  \hfill
  \begin{subfigure}[b]{0.48\columnwidth}
    \centering
    \includegraphics[width=\textwidth,height=0.8\textwidth]{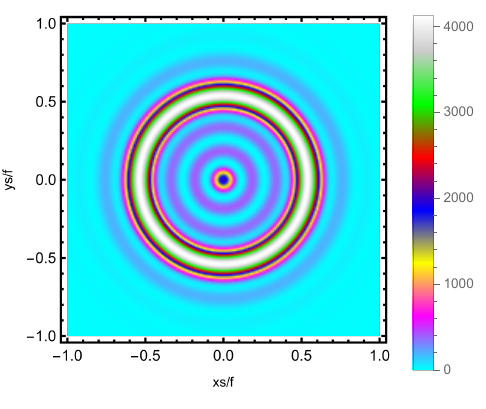}
    \caption{$e=10$}
  \end{subfigure}
\begin{subfigure}[b]{0.48\columnwidth}
    \centering
    \includegraphics[width=\textwidth,height=0.8\textwidth]{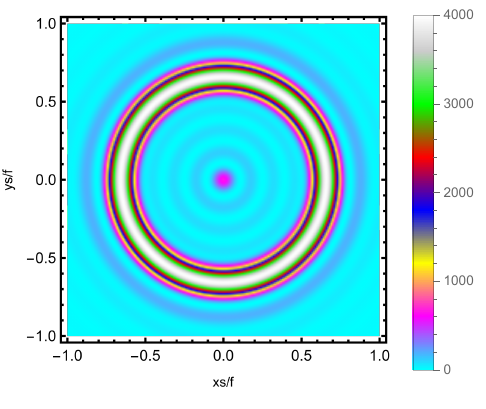}
    \caption{$e=15$}
  \end{subfigure}
  \hfill
  \begin{subfigure}[b]{0.48\columnwidth}
    \centering
    \includegraphics[height=0.8\textwidth,width=\textwidth]{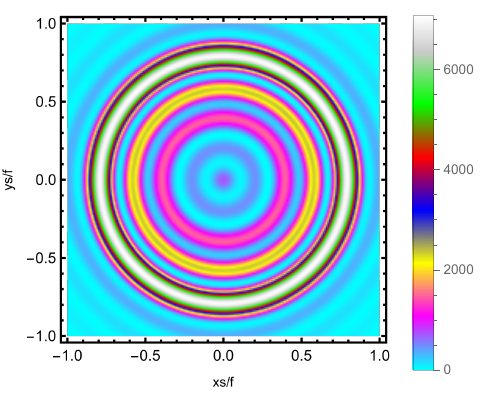}
    \caption{$e=20$}
  \end{subfigure}
  \caption{Impact of the parameter $e$ on the formation of the Einstein ring, with the parameters fixed as $a=1$ $Q=0.5$, $z_{h}=1$, $\  \omega=35$, $\rho=20$}
  \label{11}%
\end{figure}

\begin{figure}[htbp]
  \centering
  \begin{subfigure}[b]{0.48\columnwidth}
    \centering
    \includegraphics[width=\textwidth,height=0.8\textwidth]{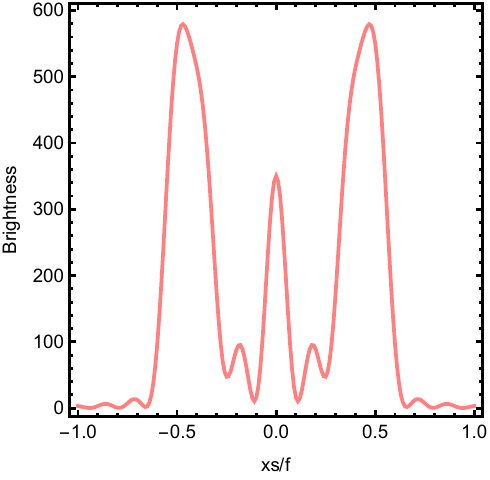}
    \caption{$e=5$}
  \end{subfigure}
  \hfill
  \begin{subfigure}[b]{0.48\columnwidth}
    \centering
    \includegraphics[width=\textwidth,height=0.8\textwidth]{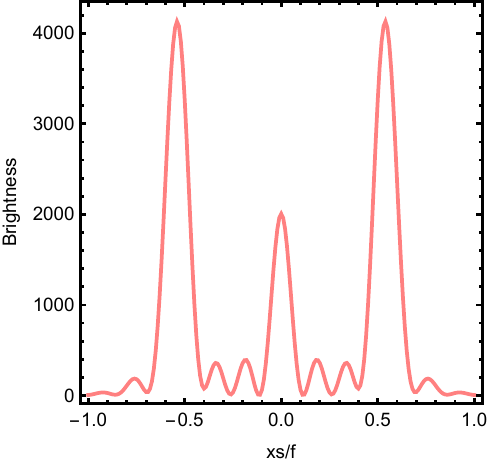}
    \caption{$e=10$}
  \end{subfigure}
\begin{subfigure}[b]{0.48\columnwidth}
    \centering
    \includegraphics[width=\textwidth,height=0.8\textwidth]{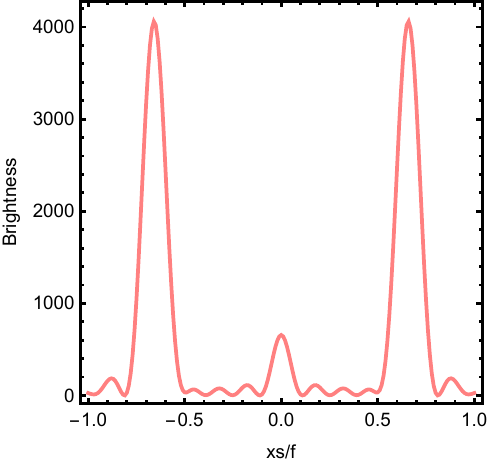}
    \caption{$e=15$}
  \end{subfigure}
  \hfill
  \begin{subfigure}[b]{0.48\columnwidth}
    \centering
    \includegraphics[height=0.8\textwidth,width=\textwidth]{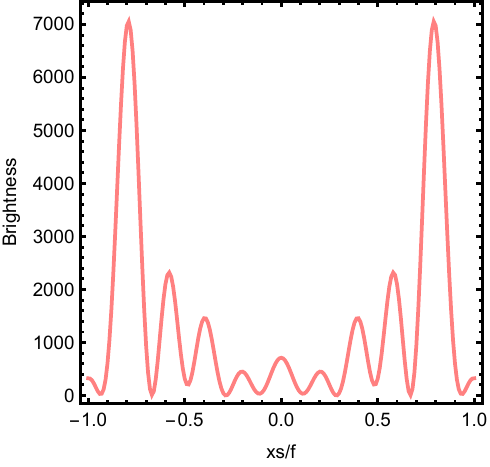}
    \caption{$e=20$}
  \end{subfigure}
  \caption{Influence of the parameter $e$ on the brightness, with the observational angle set to 
$\theta_{obs}=0$ and the remaining parameters fixed as 
$a=1$ $Q=0.5$, $z_{h}=1$, $\  \omega=35$, $\rho=20$}
  \label{12}%
\end{figure}

\begin{figure}[htbp]
  \centering
  \begin{subfigure}[b]{0.48\columnwidth}
    \centering
    \includegraphics[width=\textwidth,height=0.8\textwidth]{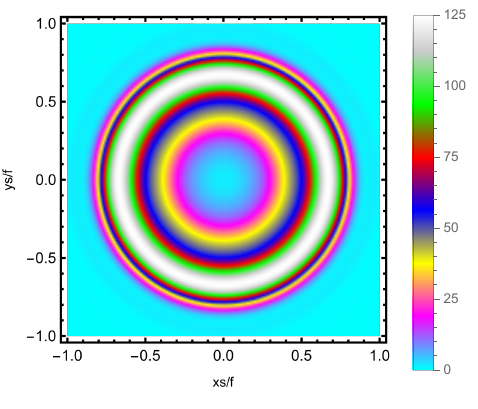}
    \caption{$\rho=10$}
  \end{subfigure}
  \hfill
  \begin{subfigure}[b]{0.48\columnwidth}
    \centering
    \includegraphics[width=\textwidth,height=0.8\textwidth]{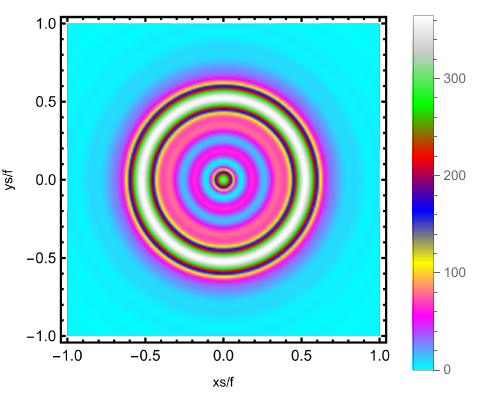}
    \caption{$\rho=15$}
  \end{subfigure}
\begin{subfigure}[b]{0.48\columnwidth}
    \centering
    \includegraphics[width=\textwidth,height=0.8\textwidth]{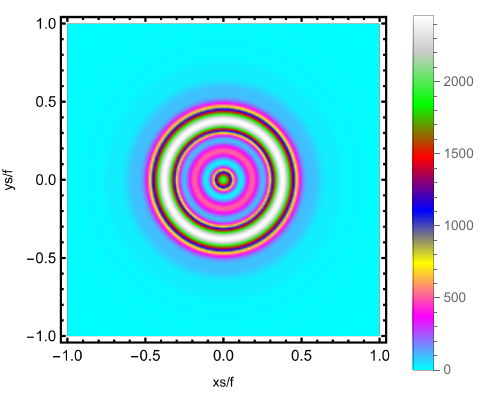}
    \caption{$\rho=20$}
  \end{subfigure}
  \hfill
  \begin{subfigure}[b]{0.48\columnwidth}
    \centering
    \includegraphics[height=0.8\textwidth,width=\textwidth]{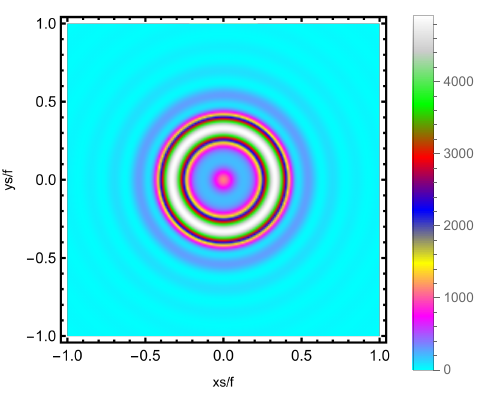}
    \caption{$\rho=25$}
  \end{subfigure}
  \caption{Effect of the parameter $\rho$ on the Einstein ring, with the observation angle fixed at 
$\theta_{obs}=0$, and the other parameters held constant as $a=1$ $Q=0.5$, $z_{h}=1$, $\omega=35$, $e=1$}
  \label{13}%
\end{figure}

\begin{figure}[htbp]
  \centering
  \begin{subfigure}[b]{0.48\columnwidth}
    \centering
    \includegraphics[width=\textwidth,height=0.8\textwidth]{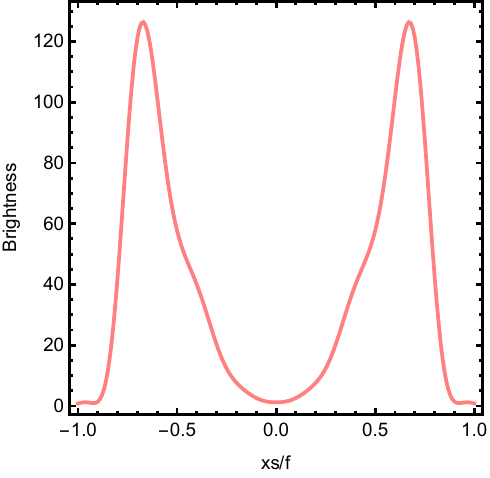}
    \caption{$\rho=10$}
  \end{subfigure}
  \hfill
  \begin{subfigure}[b]{0.48\columnwidth}
    \centering
    \includegraphics[width=\textwidth,height=0.8\textwidth]{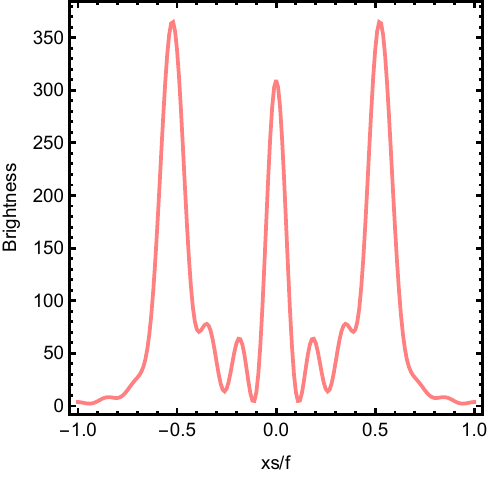}
    \caption{$\rho=15$}
  \end{subfigure}
\begin{subfigure}[b]{0.48\columnwidth}
    \centering
    \includegraphics[width=\textwidth,height=0.8\textwidth]{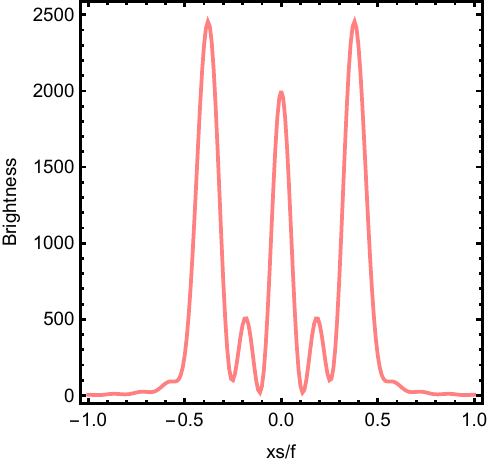}
    \caption{$\rho=20$}
  \end{subfigure}
  \hfill
  \begin{subfigure}[b]{0.48\columnwidth}
    \centering
    \includegraphics[height=0.8\textwidth,width=\textwidth]{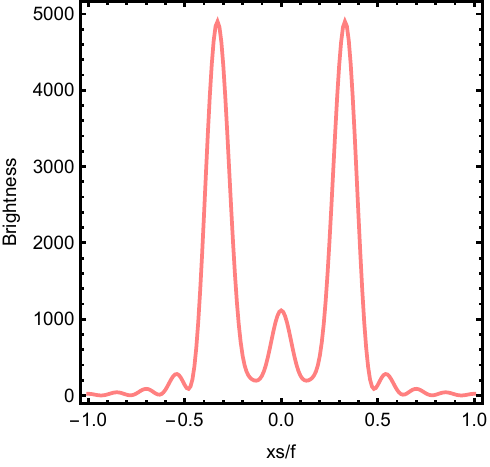}
    \caption{$\rho=25$}
  \end{subfigure}
  \caption{Influence of the parameter $\rho$ on the brightness, with the observational angle set to 
$\theta_{obs}=0$ and the remaining parameters fixed as 
$a=1$, $Q=0.5$, $z_{h}=1$, $\  \omega=35$, $e=1$}
  \label{14}%
\end{figure}

\begin{figure}[htbp]
  \centering
  \begin{subfigure}[b]{0.48\columnwidth}
    \centering
    \includegraphics[width=\textwidth,height=0.8\textwidth]{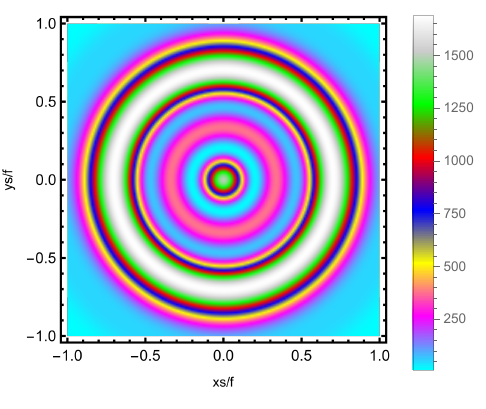}
    \caption{$\omega=20$}
  \end{subfigure}
  \hfill
  \begin{subfigure}[b]{0.48\columnwidth}
    \centering
    \includegraphics[width=\textwidth,height=0.8\textwidth]{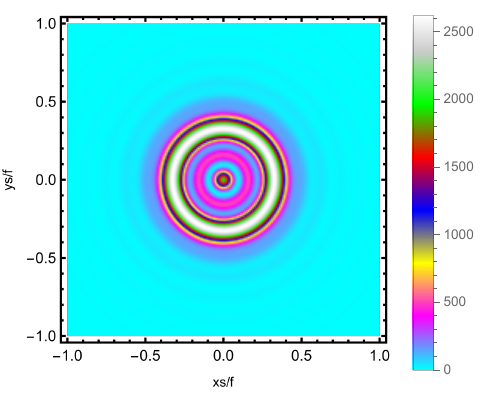}
    \caption{$\omega=40$}
  \end{subfigure}
\begin{subfigure}[b]{0.48\columnwidth}
    \centering
    \includegraphics[width=\textwidth,height=0.8\textwidth]{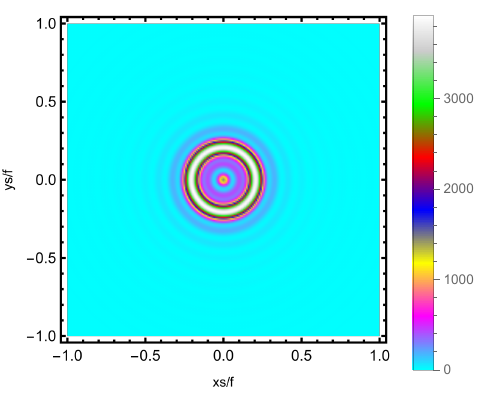}
    \caption{$\omega=60$}
  \end{subfigure}
  \hfill
  \begin{subfigure}[b]{0.48\columnwidth}
    \centering
    \includegraphics[height=0.8\textwidth,width=\textwidth]{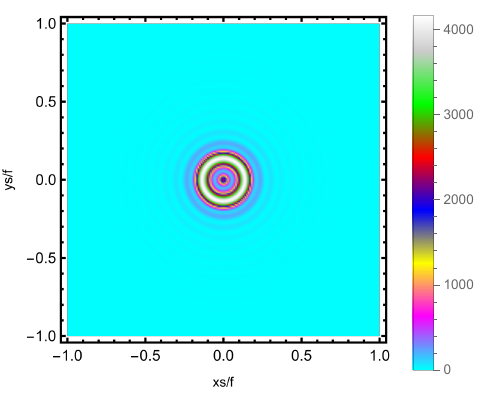}
    \caption{$\omega=80$}
  \end{subfigure}
  \caption{Effect of the parameter $\omega$ on the Einstein ring, with the observation angle fixed at 
$\theta_{obs}=0$, and the other parameters held constant as $a=1$, $Q=0.5$, $z_{h}=1$, $\rho=20$, $e=1$}
  \label{15}%
\end{figure}

\begin{figure}[htbp]
  \centering
  \begin{subfigure}[b]{0.48\columnwidth}
    \centering
    \includegraphics[width=\textwidth,height=0.8\textwidth]{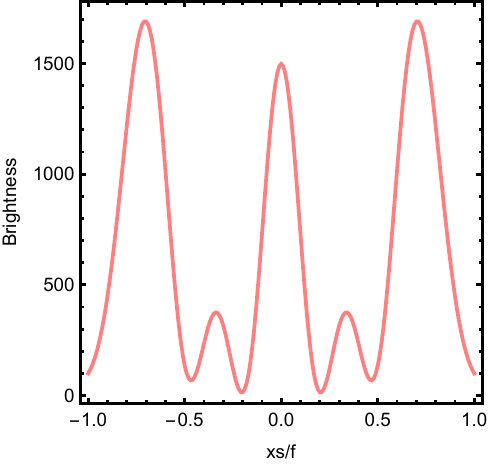}
    \caption{$\omega=20$}
  \end{subfigure}
  \hfill
  \begin{subfigure}[b]{0.48\columnwidth}
    \centering
    \includegraphics[width=\textwidth,height=0.8\textwidth]{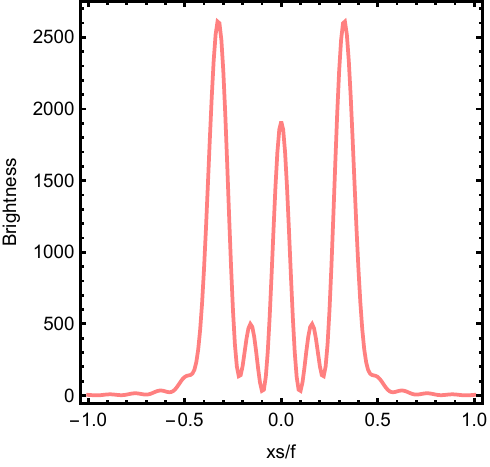}
    \caption{$\omega=40$}
  \end{subfigure}
\begin{subfigure}[b]{0.48\columnwidth}
    \centering
    \includegraphics[width=\textwidth,height=0.8\textwidth]{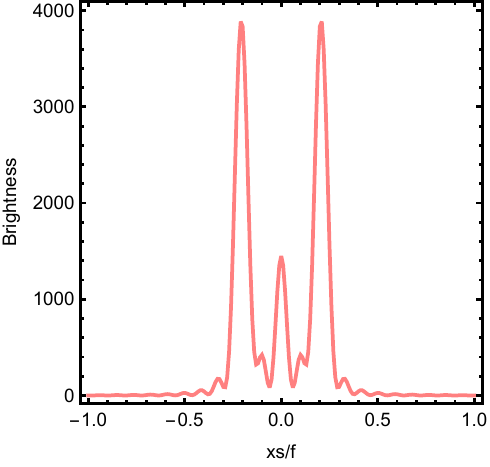}
    \caption{$\omega=60$}
  \end{subfigure}
  \hfill
  \begin{subfigure}[b]{0.48\columnwidth}
    \centering
    \includegraphics[height=0.8\textwidth,width=\textwidth]{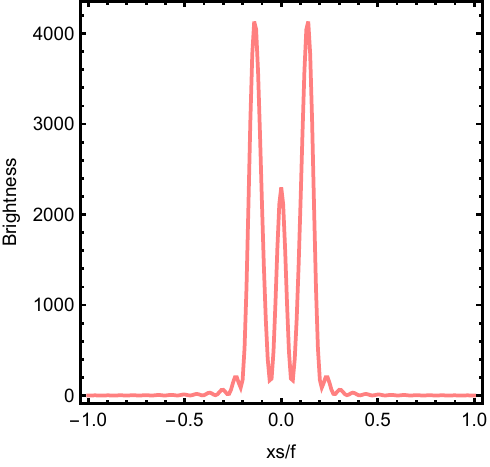}
    \caption{$\omega=80$}
  \end{subfigure}
  \caption{Influence of the parameter $\omega$ on the brightness, with the observational angle set to 
$\theta_{obs}=0$ and the remaining parameters fixed as 
$a=1$, $Q=0.5$, $z_{h}=1$, $\rho=20$, $e=1$}
  \label{16}%
\end{figure}

\begin{figure}[htbp]
  \centering
  \begin{subfigure}[b]{0.48\columnwidth}
    \centering
    \includegraphics[width=\textwidth,height=0.8\textwidth]{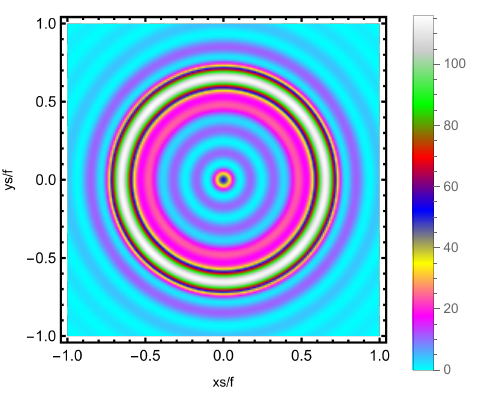}
    \caption{$T=0.352152$}
  \end{subfigure}
  \hfill
  \begin{subfigure}[b]{0.48\columnwidth}
    \centering
    \includegraphics[width=\textwidth,height=0.8\textwidth]{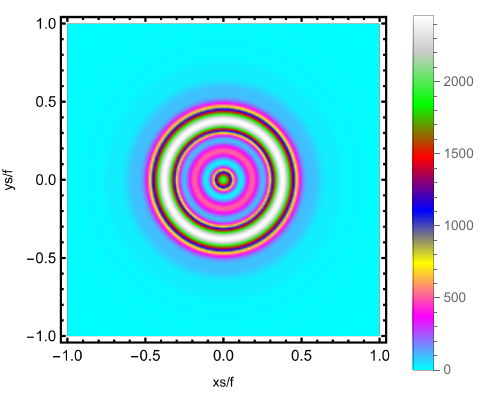}
    \caption{$T=0.299659$}
  \end{subfigure}
\begin{subfigure}[b]{0.48\columnwidth}
    \centering
    \includegraphics[width=\textwidth,height=0.8\textwidth]{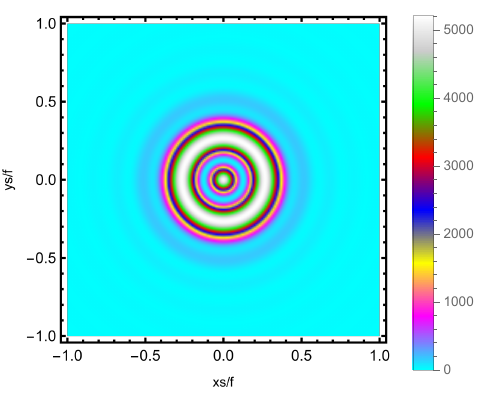}
    \caption{$T=0.264514$}
  \end{subfigure}
  \hfill
  \begin{subfigure}[b]{0.48\columnwidth}
    \centering
    \includegraphics[height=0.8\textwidth,width=\textwidth]{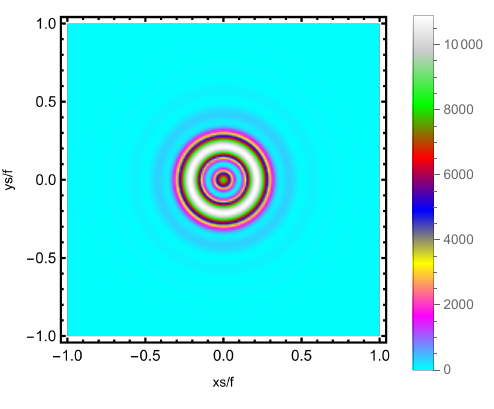}
    \caption{$T=0.240449$}
  \end{subfigure}
  \caption{Effect of temperature $T$ on the Einstein ring, with fixed parameters 
$\theta_{obs}=0$, $a=1$, $e=1$, $\rho=20$, $Q=0.5$, and $\omega=35$. Temperature decreases with increasing horizon radius $z_h=0.8,1,1.2,1.4$}
  \label{17}%
\end{figure}

\begin{figure}[htbp]
  \centering
  \begin{subfigure}[b]{0.48\columnwidth}
    \centering
    \includegraphics[width=\textwidth,height=0.8\textwidth]{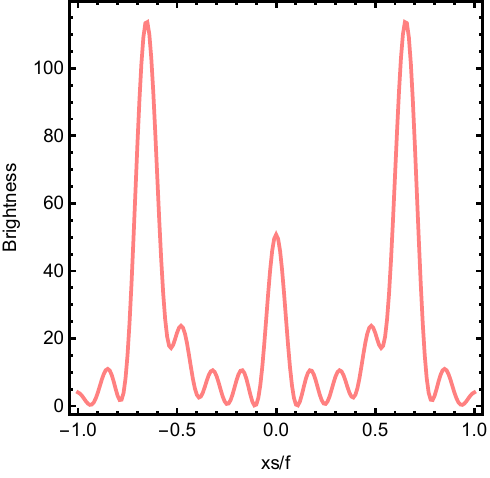}
    \caption{$T=0.352152$}
  \end{subfigure}
  \hfill
  \begin{subfigure}[b]{0.48\columnwidth}
    \centering
    \includegraphics[width=\textwidth,height=0.8\textwidth]{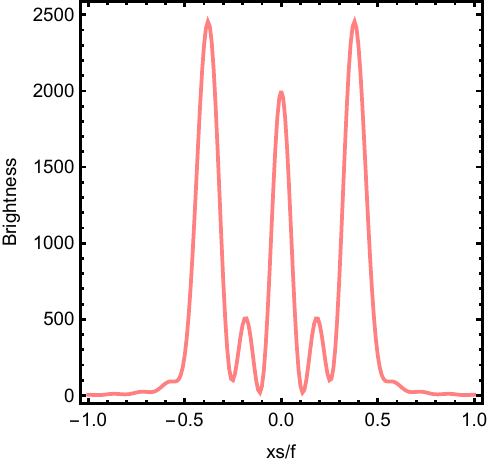}
    \caption{$T=0.299659$}
  \end{subfigure}
\begin{subfigure}[b]{0.48\columnwidth}
    \centering
    \includegraphics[width=\textwidth,height=0.8\textwidth]{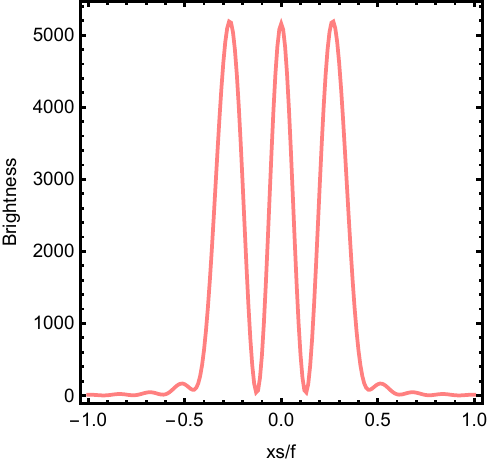}
    \caption{$T=0.264514$}
  \end{subfigure}
  \hfill
  \begin{subfigure}[b]{0.48\columnwidth}
    \centering
    \includegraphics[height=0.8\textwidth,width=\textwidth]{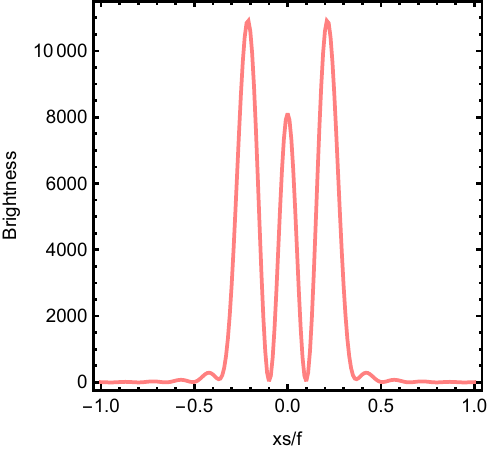}
    \caption{$T=0.240449$}
  \end{subfigure}
  \caption{Influence of the parameter $T$ on the brightness, with parameters like  $\theta_{obs}=0$, $a=1$, $e=1$, $\rho=20$, $e=0.5$, $\omega=35$ and various values of horizon radius $z_h=0.8,1,1.2,1.4$}
  \label{18}%
\end{figure}

\begin{figure}[htbp]
  \centering
  \begin{subfigure}[b]{0.48\columnwidth}
    \centering
    \includegraphics[width=\textwidth,height=0.8\textwidth]{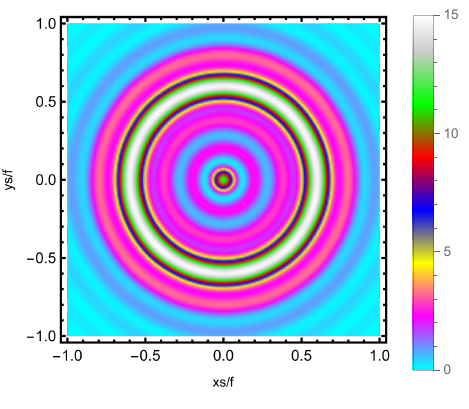}
    \caption{$\mu=0.4$}
  \end{subfigure}
  \hfill
  \begin{subfigure}[b]{0.48\columnwidth}
    \centering
    \includegraphics[width=\textwidth,height=0.8\textwidth]{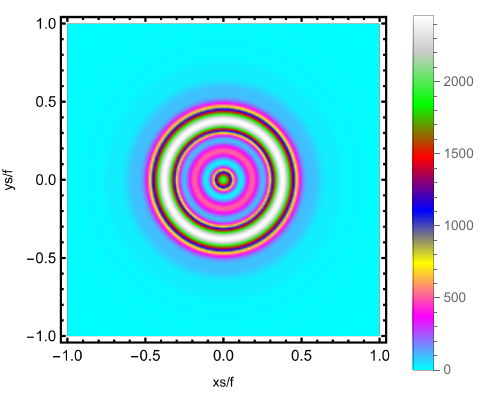}
    \caption{$\mu=0.5$}
  \end{subfigure}
\begin{subfigure}[b]{0.48\columnwidth}
    \centering
    \includegraphics[width=\textwidth,height=0.8\textwidth]{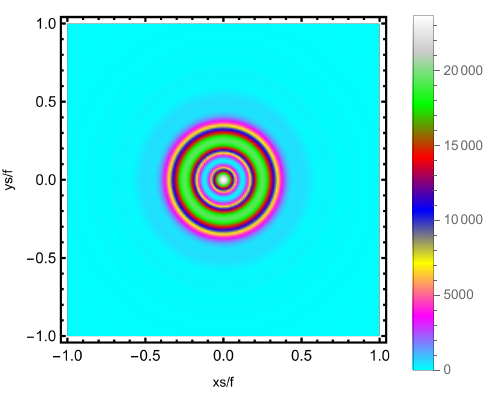}
    \caption{$\mu=0.6$}
  \end{subfigure}
  \hfill
  \begin{subfigure}[b]{0.48\columnwidth}
    \centering
    \includegraphics[height=0.8\textwidth,width=\textwidth]{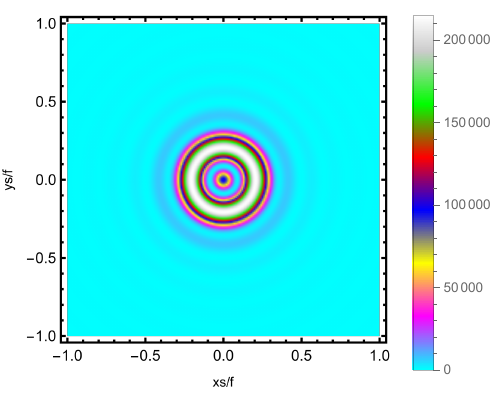}
    \caption{$\mu=0.7$}
  \end{subfigure}
  \caption{Effect of $\mu$ on the Einstein ring, where $\theta_{obs}=0$, $a=0.1$, $c=0.1$, $\Omega=-\frac{2}{3}$, $e=0.5$, $y_{h}=5$, $\omega=90$, The chemical potential vaules change from low to high, corresponding to $Q=0.0196,0.0556,0.0835,0.1022$, respectively.}
  \label{19}%
\end{figure}

\begin{figure}[htbp]
  \centering
  \begin{subfigure}[b]{0.48\columnwidth}
    \centering
    \includegraphics[width=\textwidth,height=0.8\textwidth]{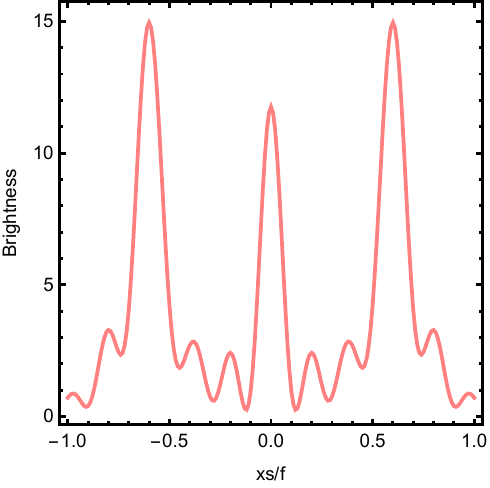}
    \caption{$\mu=0.4$}
  \end{subfigure}
  \hfill
  \begin{subfigure}[b]{0.48\columnwidth}
    \centering
    \includegraphics[width=\textwidth,height=0.8\textwidth]{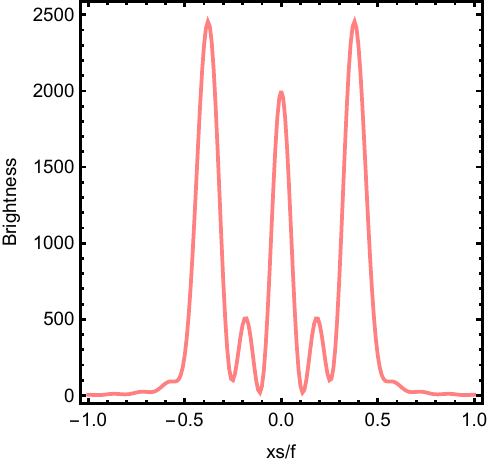}
    \caption{$\mu=0.5$}
  \end{subfigure}
\begin{subfigure}[b]{0.48\columnwidth}
    \centering
    \includegraphics[width=\textwidth,height=0.8\textwidth]{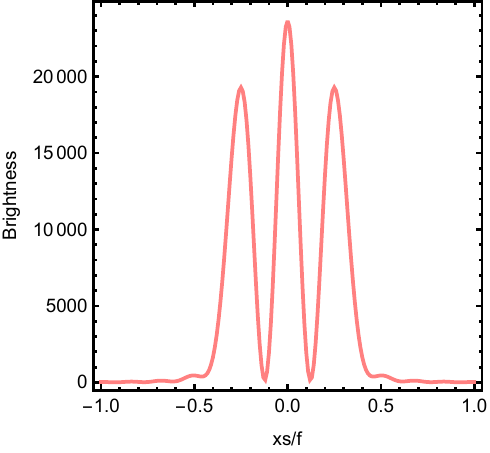}
    \caption{$\mu=0.6$}
  \end{subfigure}
  \hfill
  \begin{subfigure}[b]{0.48\columnwidth}
    \centering
    \includegraphics[height=0.8\textwidth,width=\textwidth]{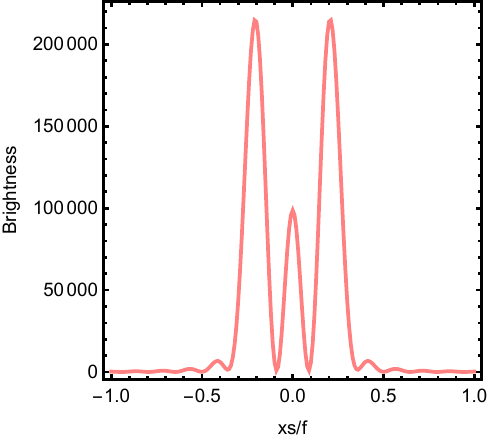}
    \caption{$\mu=0.7$}
  \end{subfigure}
  \caption{Influence on the parameter $\mu$ on the brigheness, where $\theta_{obs}=0$, $a=1$, $e=1$, $\rho=20$, $z_{h}=1$, $\Omega=35$ and different values of $Q=0.4,0.5,0.6,0.7$.}
  \label{20}%
\end{figure}

We also explore the influence of temperature $T$ on the characteristics of Einstein rings, considering a fixed chemical potential $\mu=z_h Q=0.5$. As depicted in Figure \ref{17}, a reduction in temperature value results in a gradual decrease in the Einstein ring radius. Complementarily, the corresponding luminosity distribution is shown in Figure \ref{18}, where the peak intensity is observed at $T=0.240449$ and $x_s/f=0.21$. Importantly, as the temperature progressively decreases, the x-coordinate values associated with the luminosity maxima shift towards smaller values namely $x_s/f=0.65, 0.38, 0.27, 0.21$, signifying that lower temperatures correspond to smaller Einstein ring radii.\\
The effect of the chemical potential $\mu$ on the Einstein ring is also examined, with the temperature held constant at $T=0.299659$. As shown in Figure \ref{19}, an increase in $\mu$ results in a reduction of the ring radius. This behavior is more distinctly observed in Figure \ref{20}. Specifically, for $\mu=0.4$, the peak luminosity occurs at $x_s/f=0.60$, whereas for $\mu=0.7$, the maximum shifts to $x_s/f=0.21$. These results clearly indicate that a higher chemical potential leads to a decrease in the Einstein ring radius, as evidenced by the leftward shift in the peak positions of the luminosity profile.

\section{Image formation analysed within the framework of geometric optics}

\begin{figure}
	\centering 
\includegraphics[width=0.4\textwidth, angle=0]{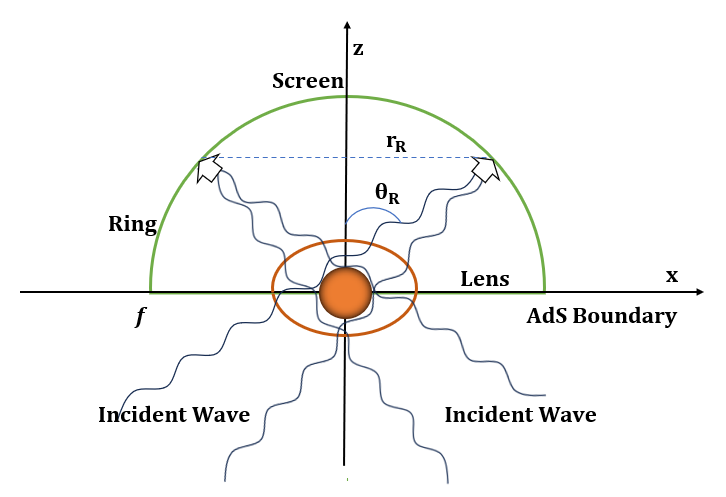}
	\caption{Correlation between the Einstein ring radius $r_R$ and the corresponding angular position $\theta_R$.} 
	\label{21}%
\end{figure}

In this section, we examine Einstein rings within the framework of geometric optics. The central objective is to investigate the photon incidence angle in the context of a spherically symmetric spacetime background, as described by Eqs.(\ref{1}) and(\ref{2}). Within this metric structure, we define the conserved quantities as $\omega*=F(r)\partial t/\partial\lambda$ and 
$L=r^2\partial \varphi/\partial t$, where $\omega*$ corresponds to the energy of the photon, $\tilde{L}$ represents its angular momentum, and $\lambda$ denotes the affine parameter along the null geodesic. Owing to the spherical symmetry of the spacetime, the analysis is restricted to photon trajectories confined to the equatorial plane, i.e., $\theta=\pi/2$. The photon's four-velocity is given by 
$\upsilon ^{\zeta } =(\mathrm{d}/\mathrm{d}\lambda )^{\zeta } $, and it satisfies the standard geodesic conditions
\begin{eqnarray}
-F(r)(\mathrm{d}t/ \mathrm{d}\lambda)^{2} +F(r)^{-1} (\mathrm{d}r/ \mathrm{d}\lambda)^{2}\nonumber\\+r^{2} \sin \theta
(\mathrm{d}\varphi / \mathrm{d}\lambda)^{2} =0,
\end{eqnarray}
 and 
 \begin{equation}
\dot{r}^{2}=\omega^{*}-\tilde{L} y(r).
\end{equation}
Here, the function $y(r)=F(r)/r^2$ and the radial component of the photon's four-velocity is given by 
 $\dot{r} =\partial r /\partial \lambda$. The incident angle is characterized by the vector 
$n^{\zeta } \equiv \partial / \partial r^{\zeta }$, which serves as the radial direction. The boundary angle $\theta_{in}$, defined with respect to the normal vector, is introduced following the approach outlined in \cite{Liu:2022cev,Zeng:2023tjb} 
 \begin{equation}
\cos \theta _{in}=\frac{g_{\alpha \beta } \upsilon ^{\alpha } n^{\beta } }{\left | \upsilon  \right | \left | n \right | }\mid _{r=\infty }
=\sqrt{\frac{\dot{r}^{2} /F }{\dot{r}^{2} /F+\tilde{L} /r^{2} } } \mid _{r=\infty },
\end{equation}
 which we can simplify to get
  \begin{equation}
\sin\theta _{in}^{2} =1-\cos \theta _{in}^{2} =\frac{\tilde{L} ^{2}y(r) }{{\dot{r}^{2}+\tilde{L} ^{2}y(r)} }\mid _{r=\infty }=\frac{\tilde{L} ^{2}}{(\omega^{*})^{2} }.
\end{equation}
When the two endpoints of a geodesic lie along the axis passing through the center of the black hole, the axial symmetry of the spacetime ensures that the observer perceives a circular image which is an Einstein ring whose angular radius corresponds to the incident angle $\theta_{in}$ \cite{Hashimoto:2018okj}. As a photon approaches the photon sphere, it reaches a critical orbit where it neither escapes to infinity nor plunges into the black hole, but instead enters a stable circular trajectory around it. At this juncture, the photon's angular momentum is denoted by $L$, and its orbital dynamics at the photon ring are governed by the following conditions \cite{Hashimoto:2019jmw,Hashimoto:2018okj,Liu:2022cev}
 \begin{equation}
\dot{r}=0,\mathrm{d}z /\mathrm{d}r=0.
\end{equation}
In this context, the incident angle satisfies the relation 
$\sin\theta_{in}=L/\omega$. As illustrated in Figure~\ref{21}, the observed angle $\sin\theta_R$ follows a corresponding relationship, confirming the consistency of this expression
  \begin{equation}
\sin \theta _{R} =r_{R} /f.
\end{equation}
The incident angle of the photon and the angular position of the photon ring both characterize the observable angle of the ring as seen by the observer. Fundamentally, these two quantities should coincide, reflecting the same geometric interpretation of the photon’s trajectory which is
\begin{equation}
r_{R} /f=L/\omega^{*}.
\label{27}
\end{equation}

\begin{figure}[htbp]
  \centering
  \begin{subfigure}[b]{0.48\columnwidth}
    \centering
    \includegraphics[width=\textwidth,height=0.9\textwidth]{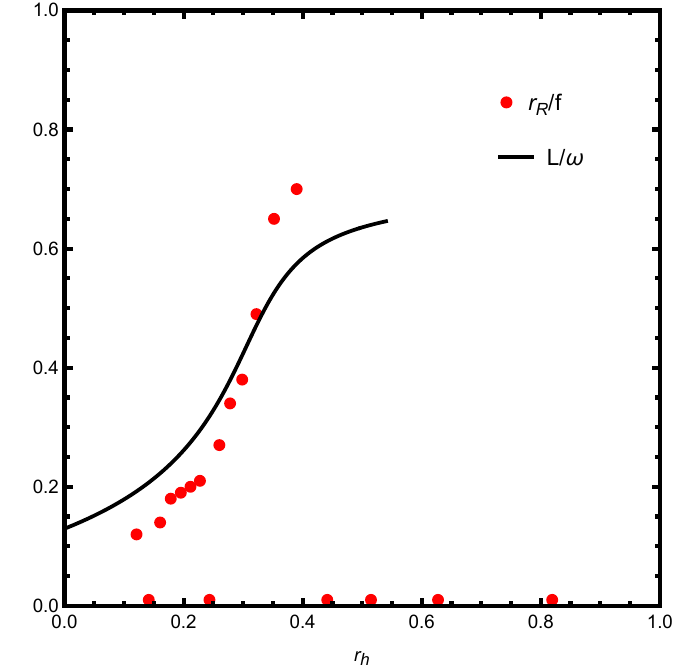}
    \caption{$a=0.01$}
  \end{subfigure}
  \hfill
\begin{subfigure}[b]{0.48\columnwidth}
    \centering
    \includegraphics[width=\textwidth,height=0.9\textwidth]{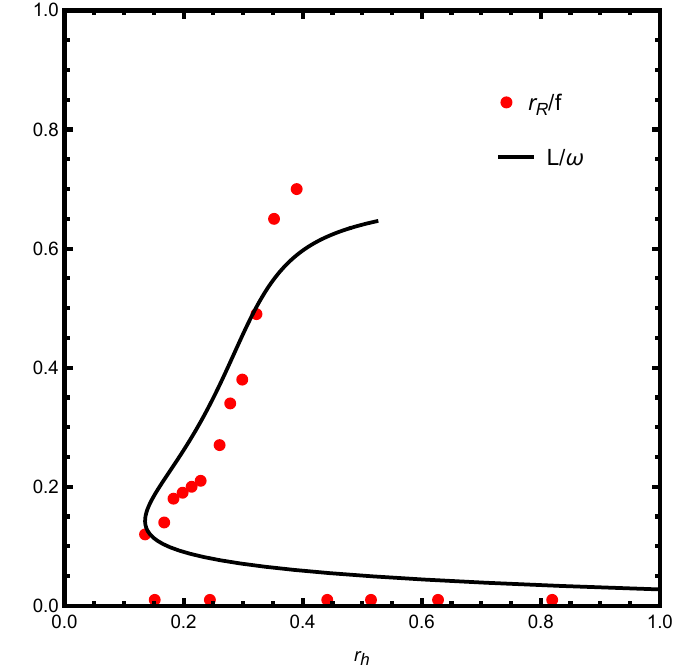}
    \caption{$a=0.1$}
  \end{subfigure}
  \hfill
  \begin{subfigure}[b]{0.48\columnwidth}
    \centering
    \includegraphics[width=\textwidth,height=0.9\textwidth]{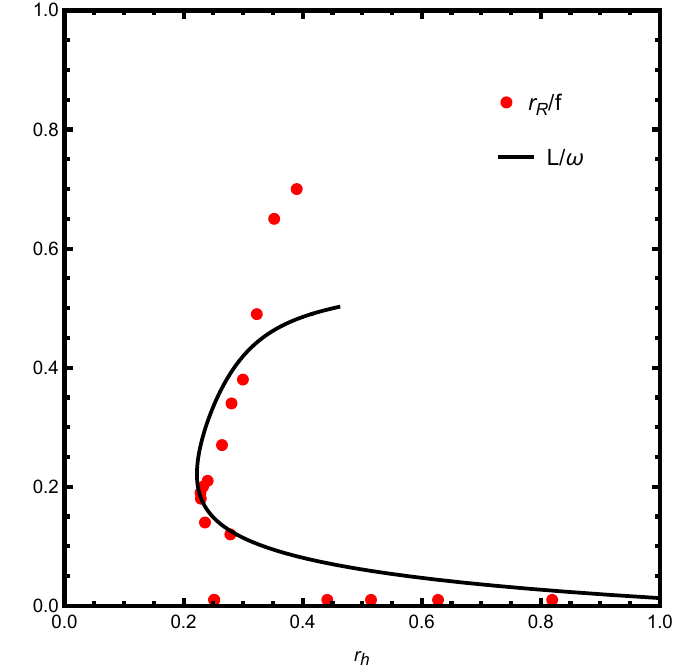}
    \caption{$a=1$}
  \end{subfigure}

  \caption{ Comparison of Einstein ring radii obtained via geometric and wave optics for varying values of the parameter $a$, with fixed values 
$\mu=0.5$, $e=1$, $\rho=20$, and $\omega=35$. The discrete red markers denote the angular positions of the Einstein rings reconstructed through wave optics, while the continuous black curve represents the photon incident angles calculated using geometric optics.}
  \label{22}%
\end{figure}

To verify the analytical expression presented in Eq.(\ref{27}), numerical methods are employed across varying physical parameters. Figure \ref{22}  illustrate the effects of the parameters $a$ on the radius of the Einstein ring, with the chemical potential fixed at $\mu=0.5$. The analysis demonstrates a strong agreement between the Einstein ring angle predicted by wave optics and the incident angle of the photon ring determined via geometric optics. This correspondence remains robust across different values of $a$, although these parameters can influence the accuracy of the numerical fit. These results collectively affirm that wave optics serves as a viable and consistent method for the holographic reconstruction of Einstein rings.

\section{Summary and conclusions}
Grounded in the framework of the AdS/CFT correspondence, this study presents a detailed investigation of the Einstein ring phenomenon arising from AdS-Resissner-Nordstr$\ddot{o}$m black holes with Euler-Heisenberg nonlinear electrodynamics, employing wave optics techniques. By examining the asymptotic behavior of the wave function near the AdS boundary, the response at the North Pole originating from a Gaussian wave source placed at the South Pole was successfully determined. The results reveal that the response function encapsulates the diffraction pattern produced by the scattering of waves off the black hole. It is observed that the amplitude of the response function diminishes with increasing values of wave source frequency $\omega$, scalar field electric charge $e$ and radial position $\rho$. Conversely, the amplitude is enhanced as the parameters chemical potential $\mu$, and temperature $T$ increase.\\
To gain deeper insight into the Einstein ring phenomenon, an optical setup incorporating a convex lens was introduced, enabling a systematic examination of how various physical parameters influence the ring structure. The analysis reveals that the radius of the Einstein ring decreases with increasing values of the parameters $\rho$, $\omega$, $T$ and $\mu$. In contrast, the ring radius expands with increasing values of the parameter $e$ and remains same for the increasing values of $a$. Moreover, the visibility and definition of the ring improve as the wave source frequency $\omega$ increases, indicating enhanced resolution at higher frequencies.\\
Additionally, within the framework of geometric optics, the incident angle of photons was thoroughly analyzed. Both theoretical considerations and numerical simulations demonstrate that the results obtained from geometric optics align closely with those derived through wave optics. This consistency remains valid across varying values of the parameters $a$, $e$, $\rho$, chemical potential 
$\mu$, and temperature $T$, although these parameters do affect the accuracy of the fitting.\\ Variations in the Euler–Heisenberg quantum correction parameter 
$a$ do not lead to appreciable changes in either the amplitude of the response function or the Einstein ring radius. This can be attributed to the fact that the quantum correction term contributes subdominantly in the near-boundary region and introduces only minor modifications to the effective potential governing photon orbits. As such, the geometrical and wave-optical features remain largely unaffected for small to moderate values of 
$a$, consistent with its role as a perturbative correction in the semi-classical regime.\\
This study also investigates the influence of the scalar field electric charge $e$ and the radial position $\rho$ of the wave source on the formation of Einstein rings, the amplitude of the response function, and the brightness peak with respect to the normalized screen coordinate $x_s/f$. To our knowledge, this aspect has not been explored in prior works. The results demonstrate that both the scalar charge and the radial emission point significantly affect the observable characteristics of the ring. The scalar field electric charge modifies the coupling between the scalar and electromagnetic fields in the bulk, thereby altering the propagation dynamics of the wave and reshaping the boundary response. An increase in scalar charge effectively changes the electromagnetic lensing potential, which in turn shifts the peak brightness and modifies the angular size and clarity of the Einstein ring. Similarly, the radial position of the source determines the effective geodesic paths and phase accumulation of the wavefronts as they traverse the curved spacetime, leading to variations in the diffraction pattern observed on the boundary. These dependencies indicate a deeper sensitivity of holographic imaging to the internal structure and field content of the bulk spacetime, suggesting that Einstein rings can serve not only as probes of gravitational geometry but also as diagnostic tools for the scalar–electromagnetic interactions and source configurations in the dual field theory.

\appendix




\end{document}